\documentclass[aps,showpacs,superscriptaddress,10pt,twocolumn,hidelinks]{{revtex4-2}}
\usepackage{amsmath}
\usepackage{amsfonts}
\usepackage{graphicx}
\usepackage{float}
\usepackage{color}
\usepackage{bm}
\usepackage{array}
\usepackage{mathtools}
\usepackage[T1]{fontenc}
\usepackage{bigints}
\usepackage{url}
\usepackage{natbib}
\usepackage{tikz}
\usetikzlibrary{matrix}
\usepackage{notoccite}

\begin{document}
	
	\title{Intermodal entanglement in a quantum optical model of HHG due to the back-action on the driving field 	}
	
	\author{Ákos Gombkötő}
	\affiliation{HUN-REN Wigner Research Centre for Physics, Konkoly-Thege M. \'{u}t 29-33, H-1121 Budapest, Hungary}
	
	\author{Peter Adam}
	\affiliation{HUN-REN Wigner Research Centre for Physics, Konkoly-Thege M. \'{u}t 29-33, H-1121 Budapest, Hungary}
	\affiliation{Institute of Physics, University of Pécs, Ifjúság útja 6, H-7624 Pécs, Hungary}
	
	\author{David Theidel}
	\affiliation{ Laboratoire d’Optique Appliquée (LOA), CNRS, École polytechnique,
		ENSTA, Institut Polytechnique de Paris, Palaiseau, France
	}
	
	\author{Tamás Kiss}
	\affiliation{HUN-REN Wigner Research Centre for Physics, Konkoly-Thege M. \'{u}t 29-33, H-1121 Budapest, Hungary}    
	
	\begin{abstract}
		
		Preparation of nonclassical light with special quantum properties is essential for quantum technologies. High-harmonic generation (HHG) is a process which not only enables the creation of attosecond pulses but also has the potential to generate light with intricate quantum properties. In a recent experiment [PRX Quantum 5. 040319], nonclassical inter-harmonic correlations have been measured from a HHG source between low-order harmonics. In this work, we theoretically investigate entanglement between different harmonics within an effective, phenomenological quantum optical model. 
		This model implements a significant degree of simplification regarding the processes within the target material, treating the material through susceptibilities, as it is usual in quantum optics.
		Such an approach yields a general description of HHG in the few-harmonic generation regime, permitting the implications that can be derived within it to hold broadly within the domain of validity.
		We find that entanglement is produced as a result of the often neglected back-action. We can qualitatively reproduce experimentally measured nonclassicalities, which suggests that intermodal entanglement can, to an extent, be considered a universal phenomenon associated with HHG, rather than a result of using specific material targets.
	\end{abstract}
	\maketitle 
	
	\section{Introduction}	
	
	Established techniques for generating quantum light at optical frequencies \cite{PhysRevA.109.040101,BARNETT201719} typically rely on nonlinear optical processes \cite{SHI2025100577}. These processes involve the nonlinear optical response of materials, often described using a perturbative approach, where the induced polarization is proportional to a low-order power of the applied electric field \cite{drummond_hillery_2014}. These models all share the common feature of incorporating the material response through parameters that characterize the effective interaction between the involved modes. The resulting Hamiltonian is typically used to describe quantum optical processes, such as second- or third-harmonic generation or parametric down-conversion.
	
	The process of high-harmonic generation (HHG) includes a very intense electromagnetic pulse generating radiation across a wide spectrum of higher frequencies \cite{FMLA88,PS89,GD11,L94}. As such, HHG is a fundamental source of ultra-short pulses in attophysics. Due to the non-perturbative, strong-field effects inherent in HHG, the detailed description is highly complicated. Historically, a quantum optical description of HHG was given for the first time by J.Bergou and S.Varró in 1981 \cite{JV99}. Since then,
	various approaches have been proposed in the literature for incorporating  quantum features of the participating fields \cite{KMOV99,GSE2000,GLFG2000,PhysRevLett.106.023001,YZSSG15,PhysRevA.89.063827,22abstractconf}.  After 2020, there has been a significant surge of interest toward this topic, marking the transition of the quantum optical investigation of HHG from a niche topic into a promising research area. For reviews of research directions, techniques and methods, see Refs.~\cite{Gombkt2021AMK,8thAttoSci,SHI2025100577,Stammer_Lewenstein_2023,doi:10.1142/9789811279560_0007,Rivera+2025+1837+1855,photonics8070263,photonics8070269,43047dc146e64dd1980cdee6d66fe5b8,PhysRevA.110.043119,PhysRevA.111.013113,PhysRevA.110.043115,11vz-9gcz,Lamprou_2025}.
	
	In recent years, both theoretical and experimental results suggest that the quantum optics of high-intensity light-matter interaction--specifically of HHG-- have remarkable potential in developing a source of unusual nonclassical photonic states. Details of experimental setups and measured results may be found in e.g. Refs.~\cite{PhysRevA.89.063827,TKDF19,TKDF19,Rasputnyi2024,PRXQuantum.5.040319,Nayak2025,Lemieux2025,6r6n-pxfp,photonics8060192,Lamprou_2025,CruzRodrguez2024QuantumPI}.
	Theoretical research typically uses models where the target material and either the driving field or the harmonic modes are quantized. This naturally leads to approximations, most often non-depletion for either the driving field, or for the initial material quantum state. 
	The theoretical calculations suggest that large-scale, multimode correlated 
	\cite{GFV21Q,GAthesis,PhysRevX.15.011023,PhysRevB.109.035203,SPRJLTL21,STPH22,Pizzi2023,PhysRevResearch.5.043138,PhysRevLett.132.143603,PhysRevB.109.125110,PhysRevA.110.063704,PhysRevA.110.023115,PhysRevA.109.033706} and squeezed \cite{JV99,photonics8070269,PhysRevLett.132.143603,PhysRevA.110.023115,PhysRevA.110.063118,AG19,GFV21Q,Gombk2023,PhysRevLett.132.143603,PhysRevX.15.011023,PhysRevA.109.033110,PhysRevResearch.6.033010,PhysRevA.110.063118} states, as well as highly displaced, intensive cat-states \cite{LCPRSKT21,RSPLTPLC21,SPRJLTL21,Stammer_2024,PhysRevB.109.035203,PhysRevLett.134.013601,PhysRevA.105.033714,PhysRevLett.134.013601,11vz-9gcz,PhysRevA.110.063704,3gjp-f7br,PRXQuantum.5.010328}, and other quantum states characterized by nonclassical photon-statistic \cite{AG19,GFV21Q,GAthesis,PhysRevA.110.023115,PhysRevA.109.033706,GIHCGS25,dPSAHRANO25,theidel2026photonantibunching} may be generated as a result of HHG. 
	The quantum properties of harmonics generated under coherent-state excitation have been theoretically studied in e.g. Refs.~\cite{GC16,GFV21Q,AG19,Gombk2023,GAthesis}, while the modifications in the photon statistics of the driving field has been explored in Refs.~\cite{GO16,TNKIGIT17,TKDF19,G20}.   
	HHG involving non-coherent state driving pulses have also been extensively studied recently \cite{PhysRevResearch.6.033079,GTBK22,TBGK22,Rasputnyi2024,Stammer2024,4hdl-bdwj,PhysRevResearch.6.L032033,PhysRevA.111.063105,4807f9d595674c25a8147aeb297df08f,PhysRevA.111.043111,1fsq-ffsv,SPRJCMLM25,PLSPLMRJ26,SMRJLMHM26}, together with the
	impact that the initial quantum state of, and specifically initial correlations in  the target material, has on the resulting photonic properties \cite{PhysRevLett.132.143603,PhysRevLett.128.047401,PhysRevA.109.063103,PhysRevA.109.012223,PhysRevA.109.033110,PhysRevA.108.053119,PhysRevA.110.063118,Hansen_2024}.
	Brief summaries of current research can be found in sources such as Refs.~\cite{FennelT21,Tzallas2023,Tzallasa2023,KDPCorkum23,Ciappina2025,refId0,CoOr26}.

	Theoretical prediction of nontrivial cross-correlations emerging between harmonics have been made in earlier works. For a monochromatically driven two-level system, harmonic-harmonic correlations have been analyzed in Refs.~\cite{GFV21Q,GAthesis}. Afterward, a calculation involving one-photon ionization and free-electron dynamics found a necessity for correlations between harmonics to arise \cite{PhysRevResearch.5.043138}.
	Furthermore, an analysis of HHG from correlated many-atom systems point to strong positive correlation emerging between harmonic photons \cite{Pizzi2023}.
	
	Questions regarding the presence of entanglement, of course, needs to be treated separately, and with care. While entanglement has been found to emerge during HHG between electron states and light \cite{PhysRevA.110.023115,PhysRevA.109.033706}, and between the excitation mode and harmonic modes \cite{PhysRevB.109.125110,PhysRevLett.132.143603,PhysRevA.110.063704}, the focal point of our current work concerns the harmonic-harmonic entanglement.
	Using rather general physical considerations, and without relying on a particular Hamiltonian, it was shown in Ref.~\cite{SPRJLTL21}, that during the process of HHG, after a conditioning, all harmonic modes become naturally entangled.
	A more detailed mathematical discussion of these results, with emphasis on the conditioning, can be found in Ref.~\cite{STPH22}. These methods have also been applied for calculations involving semiconductor targets \cite{PhysRevB.109.035203}. 
	According to these papers, the entanglement is such, that 
	measuring one mode can leave the entanglement of the other modes intact, suggesting the ability of using HHG for generating high dimensional optical cluster-type multipartite entangled states, useful for measurement-based quantum computation \cite{PhysRevLett.86.5188,PhysRevLett.95.010501}.
	On one hand, this raises the possibility of applications, and on the other hand, a solid theoretical foundation of the results, together with a limit of validity, is highly desirable.
	A different direction of research \cite{PhysRevX.15.011023} found that transitions between different laser-dressed states of
	the material system can ultimately generate similar entanglement between different harmonics. A key consideration there was the abandonment of the coherent-state-product ansatz, which we will also build upon.

	In a recent experiment \cite{PRXQuantum.5.040319}, nonclassical intermodal correlations have been reported between pairs of harmonic modes.
	The experimental measurement involved three different material systems: Si, ZnO, and GaAs. One interpretation of the emerging nonclassicalities is based on the presence of entanglement between Gaussian states \cite{6r6n-pxfp}. Nevertheless, the experimental results naturally raises the question of whether the measured nonclassical correlations between harmonic modes indeed correspond to entanglement. Further, it would be of theoretically interest and of experimental relevance to understand whether the nonclassicalities are specific to these materials, to semiconductors in general, or are broadly material-independent, universal features of HHG?
	
	In order to answer this question, we turn to a recently introduced effective model for HHG \cite{PhysRevA.109.053717}, which considers both the driving field and the harmonics as quantized objects, while making minimal assumptions about the material target. Such an approach allows for providing a qualitative answer to this question by presenting approximate, analytical results.
	We show that in a consistent treatment of the dynamics, incorporating the depletion of the excitation mode results in a natural deviation from the coherent-state product ansatz, and entanglement within the harmonics. We use our results to qualitatively interpret the experimental results in \cite{PRXQuantum.5.040319}.
	In some respect we reinforce the results in Ref.~\cite{SPRJLTL21}, that is, we show that entanglement appear as a result of the depletion in the driving field, without conditioning, and without using heuristic assumptions.
	While serving as a justification for the approach in \cite{SPRJLTL21}, our results can also be used to delineate its validity.
	In our effective model, the depletion in the driving field mode is incorporated exactly. Our approach is, in some sense, complementary to the research direction in which the quantum dynamics of materials is to be considered in an exact manner \cite{PhysRevX.15.011023,11111222}.

	Our paper is organized as follows: In section \ref{mod}, we introduce an effective quantum optical model, along with the assumptions implicit within it. In section \ref{dynequssec}, we present  the dynamical equations for the quantum state of the electromagnetic field. Analytical results obtained through a perturbative approach are given in section \ref{analyticalresults}. The interpretation of these results, applied to both an idealized perturbative and nonperturbative setup, as well as a qualitative fitting of parameters to experimental data has been given in  Sec.~\ref{interpretation}. A discussion of the limitations of our model, and of relevant experimental considerations is given in Sec.~\ref{discussoutlook}.
	A summary of our work is presented in Sec.~\ref{conclusion}.

	\section{Model}\label{mod}
	
	In any model of HHG, there are three key elements: the excitation, the material target, and the harmonic radiation. For  first-principle calculations, all three of these must be considered in a quantized framework, which, however, presents an extremely challenging problem. 
	Therefore, here we consider the material to be a classical continuum, represented solely through effective susceptibilities. In contrast, both the driving pulse and the scattered harmonics are represented as quantized modes.

	Following this approach, we consider a Hamiltonian introduced in \cite{PhysRevA.109.053717} to describe the process of HHG
	\begin{equation}\label{parHam}
		H = \hbar \omega A^\dagger A 
		+ \hbar \omega \sum_n^{M} n a_n^\dagger a_n  
		+ \hbar \sum_{n}^{M} \chi_n \left[ A^n a_n^\dagger 
		+ (A^\dagger)^na_n \right] ,
	\end{equation}
	where $A^{(\dagger)}$ are the creation/annihilation operators associated with the pump mode of frequency $\omega$, and
	$a^{(\dagger)}_n$ are the creation/annihilation operators of the harmonic mode indexed with $n$, of frequency $n\omega$.
	The $n$ indices are a subset of positive integers, specifically $n\in\{ 2,\dots M \}\subset \mathbb{N}^+$,
	and $\chi_n \in \mathbb{R}$ denotes the $n$th order susceptibility of the material. Practically $M$ corresponds to the cutoff harmonic order.

	This Hamiltonian describes  a physical process in which $n$ photons are absorbed from the driving field's mode and simultaneously a photon is emitted into the corresponding $n$th harmonic mode. The corresponding Hermitian conjugate term describes the inverse process. This interaction term is a plausible description of HHG if the electrons return to their initial state after the interaction, meaning that energy is effectively not stored in the matter, it being responsible only for mediating the interaction between modes. Consequently, effective energy transfer happens solely between the pump field and harmonic modes,  corresponding to the conservation of the energy of the electromagnetic field.
	The assumption of a single electromagnetic mode representing the excitation is commonly used in quantum optics. 
	The Hamiltonian does not include harmonic-harmonic interactions, as the harmonic modes are typically lowly populated, hence nonlinear interactions between them are weak. 
	Whenever this Hamiltonian is applied to describe a concrete realization of HHG processes, the fitting can be performed by adjusting the susceptibilities in this model. In a typical HHG process, some of the outgoing modes, e.g., the ones with even indices, are missing or negligible, which can be taken into account by choosing the corresponding susceptibilities to be zero. 
	The model of Eq.~(\ref{parHam}) opens the possibility of gaining analytical insight into the emergence of non-classicality and intermodal correlations solely due to the back-action on the driving field.
	
	The resulting zeroth-order approximate solution of the Schrödinger-equation, governed by this Hamiltonian, reads
	\begin{equation}\label{classical}
		\vert\Psi^{(0)}(t)\rangle =
		\vert \underbrace{\alpha_{0} e^{-i\omega t} }_{\alpha(t)} \rangle 
		\otimes 
		\prod_{n=2}^{M} 
		\vert \underbrace{ -it \chi_n \alpha^n_{0} e^{-in\omega t}  }_{\beta_n(t) } \rangle_{n}.
	\end{equation}
	assuming initial condition of the system
	\begin{equation}\label{initial}
		\vert \Psi(0)\rangle = \vert \alpha_{0}\rangle \otimes \vert 0 \rangle_{2} \otimes \cdots \otimes \vert 0 \rangle_{M} ~ .
	\end{equation}
	The $\vert\Psi^{(0)}(t)\rangle$ zero-order solution corresponds to the assumption of non-depleting coherent excitation with amplitude $\alpha_{0}$ and emerging harmonics in coherent states with linearly increasing amplitudes.
	This solution has been the foundation of further considerations in, for example, Refs.~\cite{LCPRSKT21,STPH22,GTBK22,PhysRevResearch.6.L032033}. It is easy to see that since the quantum state can be written as a product of coherent states, the zero-order solution does not contain entanglement or intermodal cross-correlations. However, these have been measured \cite{PRXQuantum.5.040319}, and theoretically predicted \cite{GFV21Q}, hence, it is important to investigate higher orders. 
	
	\section{Dynamical equations}\label{dynequssec}
	
	In this section, we express the dynamical equations governing the time-evolution of quantum states in a way that is conducive to further analysis. Our objective is to capture the leading deviations from the uncorrelated coherent-state product, as expressed in Eq.~(\ref{classical}), induced by the time-evolution under Hamiltonian (\ref{parHam}). For this purpose, we prepare a perturbative calculation, by an appropriate unitary transformation and by introducing a suitable interaction picture.
	
	Our aim is to use transformations that ultimately lead to the perturbative results naturally being expressed in the basis
	\begin{equation}\label{dispnumbasis}
		|\alpha(t) , J \rangle \otimes \prod_{n=2}^{M} | \beta_n(t) , j_n \rangle_n
		\hspace{1.2cm}
		J,j_n\in \mathbb{N}\,.
	\end{equation}
	Here, and later, $|x,k\rangle$ denotes the displaced number-state $\mathcal{D}(x)|k\rangle$, with the number-eigenvalues denoted with $J$ for the driving mode, and with $j_n$ for the $n$'th harmonic modes, respectively. Note that for parameters $J=0,~ \{j_n=0\}_n$, the state corresponds to Eq.~(\ref{classical}). 
	Keeping the above in mind, we apply a time-dependent unitary transformation to transform Eq.~(\ref{classical}) to the vacuum-state product 
	\begin{equation}
		|\Psi'^{(0)}(t)\rangle=e^{\Lambda(t)}|\Psi^{(0)}(t)\rangle=|0\rangle\otimes\prod_{n=2}^{M}|0 \rangle_n \, ,
	\end{equation} 
	where the $\Lambda$ operator reads
	\begin{align}\label{Lambda}
		\Lambda(t) = 
		-\bigg( \alpha_0 e^{-i\omega t} A^\dagger - \alpha^*_0 e^{i\omega t} A
		\bigg)
		\nonumber \\ 
		- \sum^{M}_{n=2} \bigg( (-it\chi_n \alpha^n_0 e^{-in\omega t}) a^\dagger_n
		-(it\chi_n \alpha^{*n}_0 e^{in\omega t}) a_n \bigg)\, .
	\end{align}
	Applying this transformation to the Schrödinger-equation governed by Hamiltonian of Eq.~(\ref{parHam}), we get the dynamical equation \begin{equation}
		i\hbar\partial_t|\Psi'(t)\rangle=H'(t)|\Psi'(t)\rangle,
	\end{equation} where the tranformed state is
	\begin{align}\label{transforedPsi}
		|\Psi'(t)\rangle 
		=
		e^{\Lambda(t)} |\Psi(t)\rangle \, ,
	\end{align}
	and the Hamiltonian reads
	\begin{align}\label{analunitarytr}
		H'(t)= 
		e^{\Lambda(t)} H  e^{-\Lambda(t)} + i\hbar  \dfrac{\partial}{\partial t} \Lambda(t).
	\end{align} 
	Using Eqs. (\ref{Lambda}) and (\ref{analunitarytr}), ignoring constant terms and terms depending only on time, the Hamiltonian $H'(t)$ can be expressed as
	\begin{align}\label{transfHam}
		H'(t) =
		\hbar \omega  A^\dagger A 
		+ \hbar \omega \sum^{M}_{n=2}  
		n  a_n^\dagger a_n  
		\nonumber \\ 
		+ \hbar \sum^{M}_{n=2}  \chi_n 
		\bigg[ a_n^\dagger 
		\sum_{k=1}^{n} \binom{n}{k} A^{k}  \alpha^{n-k}_0 e^{-i(n-k)\omega t}
		\nonumber \\ 
		+ a_n 
		\sum_{k=1}^{n} \binom{n}{k} A^{\dagger k}  \alpha^{*(n-k)}_0 e^{i(n-k)\omega t}
		\bigg]
		\nonumber \\ 
		+ \hbar \sum^{M}_{n=2}  \chi^2_n t
		\bigg[ i \alpha^{*n}_0  
		\sum_{k=0}^{n} \binom{n}{k} A^{k}  \alpha^{n-k}_0 e^{i k\omega t}
		\nonumber \\ 
		- i \alpha^n_0 
		\sum_{k=0}^{n} \binom{n}{k} A^{\dagger k}  \alpha^{*(n-k)}_0 e^{-i k\omega t}
		\bigg].
	\end{align}
	The transformed Hamiltonian in Eq.~(\ref{transfHam}) can be compartmentalized into terms with distinct physical meanings: 
	\begin{equation}
		H'(t)=H'_F + H'_{DH}(t) + H'_{D}(t) .
	\end{equation}
	The first term, $H'_F = \hbar \omega  A^\dagger A  + \hbar \omega \sum^{M}_{n=2} \!n a_n^\dagger a_n$ governs the free evolution of the driving and harmonic field modes.
	The second Hamiltonian $H'_{DH}(t)$ term, that is the third term in Eq.~(\ref{transfHam}), describes the interaction between the pump and harmonic modes, and is proportional to the first power of the susceptibilities $\chi_n$. 
	The Hamiltonian $H'_{D}(t)$, that is, the fourth term in Eq.~(\ref{transfHam}), contributes solely to the evolution of the quantum state of the excitation mode, and is proportional to the second power of susceptibilities $\chi^2_n$. 
	
	
	To calculate the dynamics of the system, it is practical to use the interaction picture, with the  interaction Hamiltonian term being
	\begin{equation}\label{whamiltonian}
		W'(t) = H'_{DH}(t)+H'_{D}(t).
	\end{equation}
	In this picture, the time-evolution of the quantum state $|\Psi'_\text{I}(t)\rangle$ is determined by the equation
	\begin{equation}\label{SchrodInt}
		i\hbar \dfrac{\partial}{\partial t} | \Psi'_\text{I} (t)\rangle
		=
		e^{i H'_F t} ~W'(t)~e^{-i H'_F t}
		| \Psi'_\text{I} (t)\rangle.
	\end{equation}
	The solution of the equation can be written in the general form
	\begin{equation}\label{psiInteraction}
		| \Psi'_\text{I} (t)\rangle
		=
		\sum_{\mathcal{F}} C_{J,j_2,\dots j_{M}}(t) 
		~e^{i H'_F t} ~ | J,j_2,\dots j_{M} \rangle ,
	\end{equation}
	where the summation is over the Fock-space under consideration, and
	$| J,j_2,\dots j_{M} \rangle =
	|J \rangle \otimes \prod^{M}_{n=2} | j_n \rangle_n $. Here $|J\rangle$ and $|j_n\rangle_n$ are the Fock-states of the driving field mode and the $n$th harmonic modes, respectively.
	After inserting Eq.~(\ref{psiInteraction}) in Eq.~(\ref{SchrodInt}), the calculations lead to the dynamical equations that govern the coefficients, and can be expressed as
	\begin{align}\label{dynequ}
		i\hbar \dot{C}_{J',j'_2,\dots j'_{M}}(t) =
		C_{J,j_2,\dots j_{M}}(t)
		e^{i\tfrac{ E' - E }{\hbar}t}
		\nonumber \\
		\times \sum_{\mathcal{F}} 
		\langle J',j'_2,\dots j'_{M} |  
		~ W'(t) ~
		| J,j_2,\dots j_{M} \rangle ,
	\end{align}
	where
	\begin{align}
		E\equiv 
		\hbar \omega \bigg[ J + \sum_{n=2}^{M}  n j_n \bigg]~,
		~~
		E'\equiv 
		\hbar \omega \bigg[ J' + \sum_{n=2}^{M}  n j'_n \bigg].
	\end{align}
	In the following sections, we will analyze these dynamical equations perturbatively. 
	
	\section{Perturbative solution}\label{analyticalresults}
	In this section, we express the leading-order perturbations in the quantum state of the electromagnetic field.
	We will consider the initial condition
	\begin{align}\label{initialcond19}
		| \Psi'_\text{I} (0)\rangle = |0\rangle \otimes |0\dots0\rangle,
	\end{align}
	corresponding to $C_{0,0\dots 0}^{(0)} = 1$ coefficient in Eq.~(\ref{dynequ}).
	In order to express the quantum state in a transparent way, it is beneficial to rearrange the summations appearing in Eq.~(\ref{transfHam}), before proceeding to solve Eq.~(\ref{dynequ}). Utilizing the identity
	\begin{equation}\label{summationarrangement}
		\sum^{M}_{n=2} \sum^n_{k=1} f(k,n)
		= \sum^{M}_{k=1} \sum^{M}_{n=\max[2,k]} f(k,n),
	\end{equation}
	one can express $W'(t)$ as 
	\begin{align}\label{drivingterm}
		W'(t)=    
		\hbar \sum^{M}_{k=1}\sum^{M}_{n=\max[2,k]}  \chi_n 
		\bigg[ 
		\binom{n}{k}  \alpha^{n-k}_0 e^{-i(n-k)\omega t} a_n^\dagger A^{k}
		\nonumber\\ +  
		\binom{n}{k}  \alpha^{*(n-k)}_0 e^{i(n-k)\omega t} a_n A^{\dagger k}
		\bigg]
		\nonumber \\ 
		+ \hbar \sum^{M}_{k=0}\sum^{M}_{n=\max[2,k]}   \chi^2_n t
		\bigg[ i \alpha^{*n}_0  
		\binom{n}{k}  \alpha^{n-k}_0 e^{i k\omega t} A^{k}
		\nonumber\\- i \alpha^n_0 
		\binom{n}{k} \alpha^{*(n-k)}_0 e^{-i k\omega t}  A^{\dagger k} 
		\bigg] ~.
	\end{align}	
	Exploiting $A|0,0\dots0\rangle =0$ and $a_n|0,0\dots0\rangle =0$ identities, it is easy to see that the only non-disappearing terms in the first-order iteration of Eq.~(\ref{dynequ}) are 
	\begin{align}
		\langle J,j_1,\dots j_{M} | ~ W'(t) ~| 0,0,\dots 0 \rangle =
		\nonumber \\ 
		i\hbar\langle J,j_1,\dots j_{M} |  \bigg[ \sum_{n=2}^{M} \! \chi^2_n t \alpha^{*n}_0   \alpha^{n}_0 | 0,0,\dots 0 \rangle
		\nonumber\\
		-\sum_{k=0}^{M}  \sum_{n=\max[2,k]}^{M} \!\!\!\!\!\!\! \chi^2_n t e^{-i k\omega t} \binom{n}{k} \alpha^n_0 \alpha^{*(n-k)}_0   \sqrt{k!}  | k,0\dots \rangle \bigg] .
	\end{align}
	It follows that only the coefficients with $J=k>0$, $j_2=\dots=j_{M}=0$ indices can be nonzero. Expressing the above equation in an equivalent, but more transparent way, we get
	\begin{align}
		\langle k,0,\dots 0 | W'(t) | 0,0,\dots 0 \rangle =
		\nonumber \\
		i\hbar \!\!\!\!\!\!\!\!\sum_{n=\max[2,k]}^{M}  \!\!\!\!\!\!\!\! \chi^2_n t \bigg[ |\alpha_0|^{2n} \delta_{k,0}
		- \alpha^n_0 
		\binom{n}{k} \sqrt{k!}  \alpha^{*(n-k)}_0 e^{-i k\omega t}\bigg].
	\end{align}
	The first-order contributions can be thus expressed by
	\begin{align}
		\dot{C}^{(1)}_{k,0,\dots 0}(t)
		=
		\!\!\!\!\!\!\!\!\sum_{n=\max[2,k]}^{M} \!\!\!\!\!\!\! \chi^2_n t \bigg[ (\delta_{k,0} - 1) \alpha^n_0 
		\binom{n}{k} \sqrt{k!}  \alpha^{*(n-k)}_0 \bigg] ,
	\end{align}
	the integration of which is straightforward, resulting in
	\begin{align}
		C^{(1)}_{k,0,\dots,0}(t) =
		-\!\!\!\!\!\!\!\!\sum_{n=\max[2,k]}^{M} \!\!\!\!\!
		\dfrac{\chi^2_n t^2 \alpha^n_0}{2} 
		\binom{n}{k} \alpha^{*(n-k)}_0 \sqrt{k!} 
	\end{align}
	for $k\geq 1$ values. Hence, in first-order iteration, the non-normalized quantum state is written as
	\begin{align}\label{firstorderresult}
		| \Psi'_{\text{I}} (t)\rangle^{(1)} = |0\rangle \otimes |0\dots\rangle 
		-
		\sum_{k=1}^{{M}} \Theta_k t^2
		|k \rangle \otimes |0\dots\rangle ,
	\end{align}
	where 
	\begin{equation}\label{thetak}
		\Theta_k =
		\!\!\!\!\!\!\!\!\!\sum^{M}_{n=\max\{ 2, k \}}
		\!\!\!\binom{n}{k} 
		\dfrac{ \chi^2_n |\alpha_0|^{2n} }{2}
		\dfrac{\sqrt{k!} }{\alpha^{*k}_0 }.
	\end{equation}
	The physical interpretation associated with the first-order iteration is the change of the quantum state of the excitation mode due to back-action described by our model. 
	The perturbative terms in the quantum state represented by Eq.~(\ref{firstorderresult}) are proportional to powers of $\chi_n t$, therefore one may consider these to be the perturbative parameters. It follows that our results are only valid for small $\chi_n t$ values.
	
	In the second-order iteration, starting from Eq.~(\ref{firstorderresult}), one must evaluate the matrix elements $\langle J,j_2,\dots j_{M} |  W'(t) | j,0,\dots 0 \rangle$ for $j\neq0$ values. Note that  the terms in Eq.~(\ref{drivingterm}) containing $a_n A^{\dagger k}$ give no contribution, since $a_n A^{\dagger k}|j,0\dots0\rangle =0$. Considering the previous procedure, it can be foreseen that the terms proportional to $\chi_n^2$ in Eq.~(\ref{drivingterm})  
	result in contributions of magnitude $\mathcal{O}(\chi_n^4 t^4)$. Consequently, up to the precision of $\mathcal{O}(\chi_n^3 t^3)$,  the only non-disappearing terms are 
	\begin{align}
		\langle J,j_2,\dots j_{M} |  W'(t) | j,0,\dots 0 \rangle =
		\nonumber \\
		\langle J,j_1,\dots | \sum^{M}_{k=1} \hbar \!\!\!\!
		\sum^{M}_{n=\max[2,k]} \!\!\!\!\!\!\!\! \chi_n 
		\binom{n}{k}  \alpha^{n-k}_0 e^{-i(n-k)\omega t}
		\nonumber\\ \times
		a_n^\dagger A^{k} | j,0,\dots \rangle  .
	\end{align}
	It follows that the only non-negligible contribution is for indices fulfilling $J=j-k$, $j_n=1$, and $\forall q\neq n : j_q=0$ relations. 
	Finally, we obtain the dynamical equations for the second-order iteration as
	\begin{align}\label{dynequ2order}
		i\hbar \dot{C}^{(2)}_{k',0\dots 1_{n'} \dots 0}
		=  \sum^{M}_{j=1} \sum^{M}_{k=1} \hbar \!\!\!\! \sum_{n=\max[2,k]}^{M} \!\!\!\! \chi_n \binom{n}{k} 
		\alpha^{n-k}_0 
		\nonumber\\ \times e^{-i(n-k)\omega t}
		\sqrt{\tfrac{j!}{j-k!}}
		\underbrace{\langle k' | j-k \rangle}_{\delta_{k',j-k}}
		\underbrace{_{n'}\!\langle 1  | 1 \rangle_n }_{\delta_{n',n}}
		\Theta_j t^2 e^{i(k'+n'-j)\omega t} .
	\end{align}
	To clarify, Eq.~(\ref{dynequ2order}) gives the time-evolution of the coefficient for the following quantum-state component: The excitation mode is in a displaced $k'$-photon state and the $n$'th harmonic mode being a displaced 1-photon state, while all other harmonics are in coherent states.
	After some simplification, one can write
	\begin{align}
		\dot{C}^{(2)}_{k',0\dots 1_{n'} \dots 0}(t)
		=  -i \sum^{M}_{k=1}  \chi_{n'} \binom{n'}{k} \alpha^{n'-k}_0 \sqrt{\tfrac{k+k'!}{k'!}}
		\Theta_{k+k'} t^2 ~,
	\end{align}
	where the expression $\binom{n'}{k}=0$ if $k>n'$.
	The solution of Eq.~(\ref{dynequ2order}) is straightforward, and we write
	\begin{align}
		C^{(2)}_{k',\dots 0,1_{n'},0 \dots}(t)
		= -i ~\Omega_{n',k'} t^3 ~,
	\end{align}
	where
	\begin{equation}\label{Omegank}
		\Omega_{n',k'} =
		\sum^{M}_{k=1}
		\chi_{n'}  \binom{n'}{k} \alpha^{n'-k}_0
		\sqrt{\dfrac{(k+k')!}{k'!}}
		\dfrac{\Theta_{k+k'}}{3} .
	\end{equation}
	The physical interpretation associated with the second-order iteration is a multitude of simultaneous modifications of the quantum state of the driving field and of one harmonic mode.
	
	\bigskip
	
	Finally, the quantum state after the second-order iteration can be written as
	\begin{align}\label{31egy}
		| \Psi'(t)\rangle^{(2)} =
		|0 \rangle 
		\!\otimes \!
		\prod_{n=2}^{M} 
		\vert 0 \rangle_{n}
		- \sum_{k=1}^{M} \Theta_k t^2
		| k \rangle 
		\!\otimes\! \prod_{n=2}^{M}
		\vert 0 \rangle_{n}  e^{-ik\omega t}
		\nonumber\\ 	
		-i	\sum_{k=0}^{M} \sum_{j=2}^{M}
		\Omega_{j,k} t^3 ~
		| k \rangle 
		\otimes \!\prod_{\substack{n=2\\ n\neq j}}^{M} \! 
		\vert 0 \rangle_{n} 
		\!\otimes\! \vert 1 \rangle_{j} 
		e^{-i(k+j)\omega t} ~.
	\end{align} 
	To illustrate the emerging entanglement between the harmonics, we specifically consider the case of $N=3$, that reads
	\begin{align}
		| \Psi'(t)\rangle = \left( |0 \rangle - \sum_{k=1}^{3} \Theta_k t^2 e^{-ik\omega t} | k \rangle  \right)\!\otimes \! \vert 0 \rangle_{2} \vert 0 \rangle_{3}
		\nonumber\\ 	
		-i \sum_{k=0}^{3}
		\Omega_{2,k} t^3 e^{-i(k+2)\omega t} ~| k \rangle 
		\otimes 
		\vert 1 \rangle_{2} \vert 0 \rangle_{3}  
		\nonumber\\ 
		-i \sum_{k=0}^{3} \Omega_{3,k} t^3 e^{-i(k+3)\omega t}~| k \rangle 
		\otimes 
		\vert 0 \rangle_{2} \vert 1 \rangle_{3} .
	\end{align}
	Here, one can observe that a Bell-state-like correlation appears between the harmonic modes.
	
	For the sake of completeness, we derive the original $|\Psi(t)\rangle$ quantum state by inverting the transformations of Eq.~(\ref{transforedPsi}) which results in
	\begin{align}\label{persol}
		| \Psi (t)\rangle =
		|\alpha_0 e^{-i\omega t} \rangle 
		\!\otimes \!
		\prod_{n=2}^{M} 
		\vert -it \chi_n \alpha^n_{0} e^{-in\omega t} \rangle_{n}
		\nonumber\\ 	-
		\sum_{k=1}^{M} \Theta_k t^2
		|\alpha_0 e^{-i\omega t}, k \rangle 
		\!\otimes\!\! \prod_{n=2}^{M} \! 
		\vert -it \chi_n \alpha^n_{0} e^{-in\omega t} \rangle_{n} e^{-ik\omega t}
		\nonumber\\ 	
		-i	\sum_{k=0}^{M} \sum_{j=n_{1}}^{M}
		\Omega_{j,k} t^3 ~
		|\alpha_0 e^{-i\omega t}, k \rangle 
		\otimes \! \prod_{\substack{n=2\\ n\neq j}}^{M} \!
		\vert \! -it \chi_n \alpha^n_{0} e^{-in\omega t} \rangle_{n} 
		\nonumber\\ 
		\otimes \vert \!-it \chi_j \alpha^j_{0} e^{-ij\omega t}, 1 \rangle_{j}  e^{-i(k+j)\omega t}
		~.
	\end{align}
	Accordingly, the resulting quantum state of the exciting and harmonic modes is an entangled superposition state in the displaced number-state basis. 
	Based on these considerations, it is clear that second-order iteration is needed to reveal intermodal correlations.
	The quantum state of the harmonic field within Eq.~(\ref{persol}) can be qualitatively described as a displaced generalized W-like state. We note that approximate W-states have been predicted to occur, using a different approach, in Ref.~\cite{STPH22} for low harmonic conversion efficiencies. We note that W-states are generally considered to be robust against photon-number losses \cite{PhysRevA.62.062314}. While the mode-dependent displacements can cause deviations from the quantum properties of the W-states, it seems plausible that the displacement does not deteriorate the robustness.

\begin{figure}[h!]
		\centering
		\includegraphics[width=0.98\linewidth]{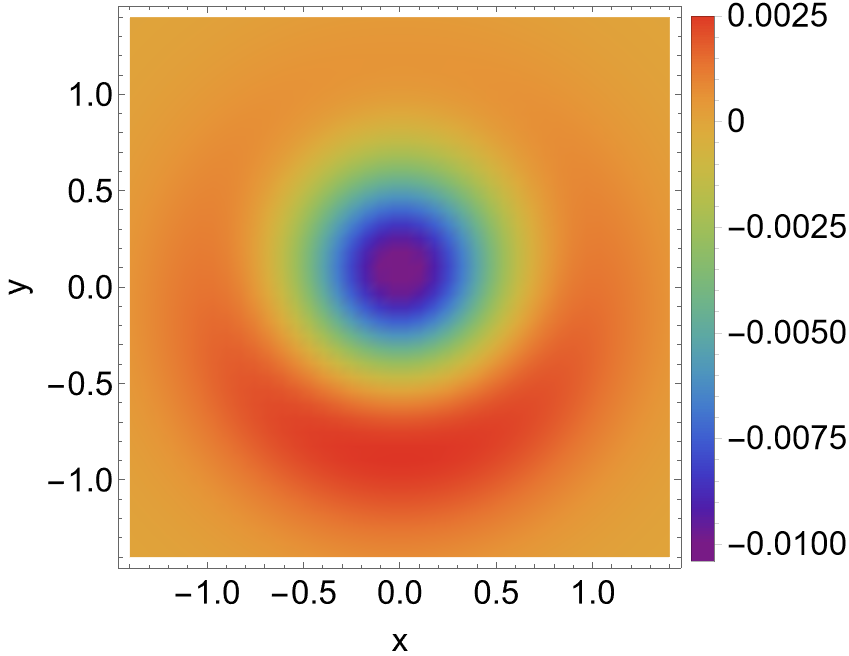}
		\caption{Deviation in the Wigner function of the 10'th harmonic, $\tfrac{\rho'^{(2)}_{10}(t)}{\text{Tr}[\rho'^{(2)}_{10}(t)]}-|0\rangle_{10}\langle0|_{10}$, based on [Eq.~(\ref{densitm})], for driving pulse of 2 optical cycles. Parameters are chosen as $|\alpha_0|=10^2$, $t=2\tfrac{2\pi}{\omega}$ and $\chi_n=\tfrac{1}{\sqrt{n}|\alpha_0|^n}$, with harmonic orders  $n\in{3,...10}$.}
		\label{fig:vazlat6}
	\end{figure}

	Note that both the harmonics and the driving field become slightly non-Gaussian as a result of the interaction. 
	To show the deviation from coherent states, we calculate the leading term difference arising in the Wigner-function \cite{Bartlett_Moyal_1949} of the density-matrix associated with the $n$th harmonic mode. After construction of the density matrix [see App.\ref{densityM} for details] and tracing out all but one harmonic mode, we get 
	\begin{align}\label{densitm}
		\rho'^{(2)}_{\text{n'th}}(t) = 
		\sum_{k=0}^{M} |\Omega_{n,k}|^2 t^6 
		~ |1 \rangle_{n}\langle 1|_{n}
		\nonumber\\ +
		\bigg[1 + \sum_{k=1}^{M} |\Theta_k|^2 t^4 
		+
		\sum_{k=0}^{M} \sum_{j=n_{1},j\neq n}^{M} |\Omega_{j,k}|^2 t^6 
		\bigg]
		|0 \rangle_{n}\langle 0|_n
		\nonumber\\ + 
		it^3 \bigg[ \Omega^*_{n,0} e^{in\omega t}
		|0\rangle_n \langle1|_n 
		- \Omega_{n,0} e^{-i n\omega t}
		|1\rangle_n \langle0|_n \bigg]  
		\nonumber \\
		-i t^5 \sum_{k=1}^{M} \bigg[ \Theta_k\Omega^*_{n,k} e^{in\omega t} 
		~ |0 \rangle_{n}\langle 1|_{n} 
		- \Theta^*_k\Omega_{n,k} e^{-in\omega t} 
		~ |1 \rangle_{n}\langle 0|_{n} \bigg].
	\end{align}
	We have calculated the deviation from coherent states through  $\tfrac{\rho'^{(2)}_{n'th}(t)}{\text{Tr}[\rho'^{(2)}_{n'th}(t)]}-|0\rangle_n\langle0|_n$, which is shown on Fig.~\ref{fig:vazlat6}. 
	\\
	Note that the unitary transformation connecting $\rho$ and $\rho'$ is merely a displacement. The deviation is small, and would not be  visible on plots of $\tfrac{\rho'^{(2)}_{n'th}(t)}{\text{Tr}[\rho'^{(2)}_{n'th}(t)]}$ itself.
	
	\subsection{Quantum optical quantities} \label{comparison}
	Next, we calculate the quantities that we intend to use for analysis in the next section. Using the quantum state given by Eq.~(\ref{persol}), we present the full analytical formulae in App.\ref{quantities}. Here, without loss of generality, we use the special initial condition $\alpha_0=-i|\alpha_0|$, which simplifies  the formulae. With this, we get
	\begin{align}\label{theta}
		\Theta_{k}=(-i)^k|\Theta_{k}|
		\nonumber	\\ 	
		|\Theta_{k}|=
		\!\!\!\!\!\!\!\!\!\sum^{M}_{n=\max\{ 2, k \}}
		\!\!\!\binom{n}{k} \tfrac{\sqrt{k!}}{2}
		\chi^2_n |\alpha_0|^{2n-k},
	\end{align} \begin{align}
		\Omega_{n',k'}=(-i)^{n'+k'}|\Omega_{n',k'}|
		\nonumber	\\
		|\Omega_{n',k'}|=
		\sum^{M}_{k=1}
		\tfrac{|\chi_{n'}|}{3}  \binom{n'}{k} \sqrt{\frac{(k+k')!}{k'!}}
		|\alpha_0|^{n'-k}
		|\Theta_{k+k'}|~ . \label{omega}
	\end{align}
	With these parameters, the quantum state can be evaluated, and calculations of one-time operator mean-values can be done in a lengthy, but straightforward manner.
	
	Recall that $|\Psi(t)\rangle$ is a perturbatively calculated state, accordingly, it is not normalized. For the purpose of expressing any physical quantities, first we need to calculate the square of the norm $\mathcal{N}(t)=\langle\Psi(t)|\Psi(t)\rangle$, that is  
	\begin{equation}\label{norm}
		\mathcal{N}(t) =
		1 + \sum_{k=1}^{M} |\Theta_k|^2 t^4
		+ \sum_{k=0}^{M} \sum_{j=2}^{M} |\Omega_{j,k}|^2 t^6.
	\end{equation}
	One of the most fundamental quantity for our analysis is the mean photon-number of the modes. For the excitation mode, it is $\langle N_1 (t)\rangle = \frac{1}{\mathcal{N}(t)}\langle \Psi(t)| A^\dagger A |\Psi(t)\rangle$, and for the $n$th harmonic mode, the mean photon-number is $\langle N_n (t)\rangle = \frac{1}{\mathcal{N}(t)} \langle \Psi(t)| a^\dagger_n(t) a_n(t)|\Psi(t)\rangle$, respectively:
	\begin{align}\label{drivingphotonnum}
		\langle N_1 (t) \rangle = 
		|\alpha_0|^2 + 
		\dfrac{1}{\mathcal{N}(t)} 
		\bigg[ 
		- 2 |\Theta_1||\alpha_0| t^2 
		\nonumber \\ 
		+\sum_{k=1}^{M} \bigg( k|\Theta_k|^2 + 2\sqrt{k+1} |\alpha_{0}| |\Theta_{k}||\Theta_{k+1}| \bigg) t^4 +
		\nonumber \\
		\sum_{k=0}^{M} \sum_{j=2}^{M} \bigg( k|\Omega_{j,k}|^2 +
		2\sqrt{k+1} |\alpha_{0}| |\Omega_{j,k}| |\Omega_{j,k+1}| \bigg) t^6
		\bigg],
	\end{align}	
	\begin{align}\label{meanphotonnharm}
		\langle N_n (t) \rangle = 
		\chi^2_n|\alpha_{0}|^{2n} t^2
		+ \dfrac{1}{\mathcal{N}(t)} 
		\bigg[ 
		2\chi_n |\alpha_{0}|^n |\Omega_{n,0}|  t^4
		\nonumber \\
		+ \bigg( \sum^{M}_{k=0} |\Omega_{n,k}|^2
		- 2\chi_n \sum_{k=1}^{M} |\alpha_{0}|^n |\Theta_{k}| |\Omega_{n,k}| \bigg) t^6
		\bigg] .
	\end{align}
	To analyze quantum properties of any particular mode, we will utilize the second-order one-time photon coherence functions $G_{nm}$, particularly as part of other, composite quantities.
	To characterize the driving field mode, we use 
	$G_{11}(t)= \tfrac{1}{\mathcal{N}(t)}\langle \Psi(t) | A^\dagger A^\dagger A A | \Psi(t)\rangle$, which reads as
	\begin{align}
		G_{11}(t) = 
		|\alpha_0|^4
		\!+
		\frac{1}{\mathcal{N}(t)} 
		\bigg[
		\!- 2\sqrt{2} |\alpha_0|^2 \big( |\Theta_2|
		+ \sqrt{2}|\alpha_0| |\Theta_1| \big) t^2
		\nonumber \\ 
		+\sum^{M}_{k=1} \!|\Theta_k|\! \bigg(\! |\Theta_k| \big( k(k\!-\!1\!+ \!4|\alpha_0|^2) \big) 
		+4|\alpha_0|^3\sqrt{k+1} |\Theta_{k+1}| \!\bigg)t^4
		\nonumber \\ 
		+ 4\sum^{M}_{k=0}k\sqrt{k+1} |\alpha_{0}| |\Theta_{k}||\Theta_{k+1}| t^4
		\nonumber \\ 
		+ 2\sum^{M}_{k=1}\sqrt{k+1}\sqrt{k+2} |\alpha_{0}|^2 |\Theta_{k}| |\Theta_{k+2}| t^4
		\nonumber \\
		+ \sum^{M}_{k=0}\sum^{M}_{j=2}|\Omega_{j,k}|^2 
		\big[ k(k-1) + 4|\alpha_0|^2 k\big] t^6 
		\nonumber \\ 
		+ 4|\alpha_0|^3 \sum_{k=0}^{M} \sum_{j=2}^{M}
		\sqrt{k+1} |\Omega_{j,k}| |\Omega_{j,k+1}|  t^6
		\nonumber \\
		+ 2\sum_{k=0}^{M} \sum_{j=2}^{M}
		\sqrt{k+1}\sqrt{k+2} |\alpha_{0}|^2 |\Omega_{j,k}||\Omega_{j,k+2}| t^6
		\nonumber \\ 
		+ 4 \sum_{k=0}^{M} \sum_{j=2}^{M}
		k\sqrt{k+1} |\alpha_{0}| |\Omega_{j,k+1}| |\Omega_{j,k}| t^6
		\bigg],
	\end{align}
	while for the $n$th ($n\neq 1$) harmonic mode, we use $G_{nn}(t)=\tfrac{1}{\mathcal{N}(t)} \langle\Psi(t)| a^\dagger_n a^\dagger_n a_na_n|\Psi(t)\rangle$, which we express as
	\begin{align}\label{gnn}
		G_{nn}(t)\approx 
		\chi^4_n |\alpha_0|^{4n} t^4 
		+
		\dfrac{1}{\mathcal{N}(t)} \bigg[
		4\chi^3_n |\alpha_0|^{3n} |\Omega_{n,0}| t^6 +
		\nonumber \\
		\bigg( 4\chi^2_n |\alpha_0|^{2n} \sum^{M}_{k=0} |\Omega_{n,k}|^2
		- 4\chi^3_n |\alpha_0|^{3n} \sum^{M}_{k=1} |\Omega_{n,k}||\Theta_k|  \bigg)t^8
		\bigg].
	\end{align} 
	With these quantities given, we can proceed to calculate the second-order normalized photon correlation functions $\gamma_{nm}$, which are often the primary experimentally measured quantities involving nonclassical photon states.
	The correlations of the excitation and harmonic modes are $\gamma_{11}(t) = \frac{G_{11}(t)}{\langle N_1 (t) \rangle \langle N_1 (t) \rangle}$
	and $\gamma_{nn}(t) = \frac{G_{nn}(t)}{\langle N_n (t) \rangle \langle N_n (t) \rangle}$, respectively. 
	These autocorrelation functions can be used to characterize the photon statistics of single modes, e.g. revealing super- or sub-Poissonian distributions.
	
	Turning our attention to the rich structure inherent in multimode states, the fundamental tools we use to gain insight are the second-order one-time intermodal cross-coherence functions. The coherence between the pump mode and the $n$th ($n\neq 1$) harmonic mode is $G_{1n}(t)= \tfrac{1}{\mathcal{N}(t)}\langle \Psi(t)| A^\dagger a^\dagger_n a_n A |\Psi(t)\rangle$, written as
	\begin{align}
		G_{1n}(t)\approx 
		\chi^2_n |\alpha_{0}|^{2n+2} t^2 +
		\dfrac{1}{\mathcal{N}(t)} \bigg[
		- 2\chi^2_n |\alpha_0|^{2n+1} |\Theta_{1}| t^4
		\nonumber \\ +
		2\chi_n |\alpha_0|^{n+2}  
		\bigg( |\Omega_{n,0}| t^4
		- \sum_{k=1}^{M} |\Theta_{k}||\Omega_{n,k}| t^6 \bigg) 
		\nonumber \\
		+ 2\chi_n |\alpha_0|^{(n+1)} \bigg( |\Omega_{n,1}| t^4
		- \sum_{k=2}^{M} \sqrt{k} |\Theta_{k-1}||\Omega_{n,k}| t^6 \bigg) 
		\nonumber \\+
		\sum_{k=0}^{M} |\Omega_{n,k}|^2 \big( k + |\alpha_0|^2 \big) t^6 
		-2\chi_n |\alpha_0|^n  \sum_{k=1}^{M} k |\Theta_{k}| |\Omega_{n,k}| t^6
		\nonumber \\+ 2\sum_{k=0}^{M} \sqrt{k+1} |\alpha_0||\Omega_{n,k}||\Omega_{n,k+1}| t^6
		\nonumber \\
		+ \chi^2_n |\alpha_0|^{2n}
		\bigg( \sum_{k=1}^{M} k|\Theta_{k}|^2 t^6
		+ \sum_{k=0}^{M} \sum_{j=2}^{M} k |\Omega_{j,k}|^2 t^8 \bigg)
		\nonumber \\- 2\chi_n |\alpha_0|^{n+1}  
		\sum_{k=0}^{M} \sqrt{k+1} |\Theta_{k+1}| |\Omega_{n,k}|  t^6
		\nonumber \\
		+ 2\chi^2_n |\alpha_0|^{2n+1} \bigg( 
		\sum_{k=1}^{M} \sqrt{k+1} |\Theta_{k}||\Theta_{k+1}| t^6
		\nonumber \\+ \sum_{k=0}^{M}\sum_{j=2}^{M} \sqrt{k+1} |\Omega_{j,k}||\Omega_{j,k+1}| t^8
		\bigg)
		\bigg] ~,
	\end{align}
	while between the $n$th ($n\neq 1$) and $m$th ($m\neq n$) harmonic modes, it is
	$G_{nm}(t)= \tfrac{1}{\mathcal{N}(t)}\langle \Psi(t)|a^\dagger_m a^\dagger_n a_n a_m |\Psi(t)\rangle$, which reads
	\begin{align}
		G_{nm}(t)\approx 
		\chi_n^2 \chi_m^2 |\alpha_0|^{2n+2m} t^4 +
		\nonumber \\
		\tfrac{1}{\mathcal{N}(t)} \bigg[ 
		2\bigg(\chi_n \chi^2_m |\alpha_0|^{2m+n} |\Omega_{n,0}|
		+
		\chi^2_n \chi_m |\alpha_0|^{2n+m} |\Omega_{m,0}| \bigg) t^6 
		\nonumber \\
		+ 2\chi_n \chi_m |\alpha_0|^{n+m} \sum_{k=0}^{M} 
		|\Omega_{m,k}||\Omega_{n,k}| t^8
		\nonumber \\
		+\sum_{k=0}^{M}
		\bigg( |\Omega_{n,k}|^2 \chi^2_m |\alpha_0|^{2m} 
		+ |\Omega_{m,k}|^2 \chi^2_n |\alpha_0|^{2n} \bigg) t^8 
		\nonumber \\ 
		-	2\chi_n \chi^2_m |\alpha_0|^{2m+n} \sum_{k=1}^{M} 
		|\Omega_{n,k}| |\Theta_{k}| t^8
		\nonumber \\ -
		2\chi_m \chi^2_n |\alpha_0|^{2n+m} \sum_{k=1}^{M} 
		|\Omega_{m,k}| |\Theta_{k}| t^8
		\bigg] .
	\end{align}
	From these, we can construct the experimentally measurable intermodal photon correlation functions. The correlation between the excitation and $n$th harmonic modes is $\gamma_{1n}(t) = \frac{G_{1n}(t)}{\langle N_1(t) \rangle \langle N_n(t) \rangle}$, while between $n$th and $m$th order harmonics is $\gamma_{nm}(t) = \frac{G_{nm}(t)}{\langle N_n (t)\rangle \langle N_m (t)\rangle}$, respectively.
	Of special interest to us is the $R(t)$ function defined as
	\begin{align}\label{Rdefinition}
		R(t) =
		\dfrac{\gamma^2_{nm}(t)}{\gamma_{nn}(t)\gamma_{mm}(t)}
		= \dfrac{G^2_{nm}(t)}{G_{nn}(t)G_{mm}(t)},
	\end{align}
	which have been measured in Ref.~\cite{PRXQuantum.5.040319}, and found to have values $R>1$. This violates the CBS-inequality ({C}auchy-{B}unyakovsky-{S}chwarz inequality), and hence implies nonclassical correlation in the quantum state of the considered modes \cite{PhysRevA.90.033616,PhysRevLett.108.260401}. One relevant theoretical calculation, involving atomic HHG, can be found in Ref.~\cite{dPSAHRANO25}.
	
	We are especially interested in the presence of entanglement, which, of course, is a nonclassicality itself, but is not neccesserily implied by $R>1$ relation. 	
	To measure intermodal entanglement between harmonics, we use the logarithmic negativity, applied to $\rho'^{(2)}_{nm}$ density-matrix, wherein we traced out all but two modes, corresponding to the $n$th and $m$th harmonics.
	\begin{equation}\label{lognegativit}
		E_{nm}(t)= \log_2 \left|\left| \left(
		\frac{1}{\text{Tr}[\rho'^{(2)}_{nm}(t)]}
		\rho'^{(2)}_{nm} (t) \right)^{\Gamma_n} \right|\right|_1
	\end{equation}
	When calculating Eq.~(\ref{lognegativit}), we use $(\rho'^{(2)}_{nm})^{\Gamma_n}$ derived in App.\ref{densityM} with the result expressed in Eq.~(\ref{lognegmatr}), specifically.

	\section{Interpretation of the results}\label{interpretation}
	
	In this section, we analyze how our model can account for experimentally measurable effects. The considerations laid down in the previous section leave the amplitude $|\alpha_0|$, the interaction time, or temporal pulse duration, $t=\tau$, the cutoff-order $M$, and susceptibilities $\{ \chi_n \}$ as free parameters.
	For modeling a given experiment, the parameters $(|\alpha_0|, \tau, M)$ can be chosen according to the experimentally measured values. 
	
	Since the choice of these parameters determine the validity of the perturbative calculations, we briefly discuss this issue. 
		While a rigorous evaluation of the error in the perturbative result can generally be challenging, standard practice relies on estimating the magnitude of the leading omitted higher-order term at the truncation point, alongside the ratio of successive terms. An error estimate is obtainable, provided that the successive terms decrease quickly enough to ensure convergence. In App.~\ref{addendumvalidity} we derived the condition $M^{3/2} \chi t |\alpha_0|^M<1$ under which the perturbative expansion can be considered valid. Using the the zeroth-order approximation of $\langle N_h \rangle$, presented in Eq.~(\ref{meanphotonnharm}), this condition can be written as $M^{3/2} \langle N_h \rangle^{1/2}<1$. Here $M$ is the number of significantly populated harmonic modes, and $\langle N_h\rangle$ represents the associated photon number expectation value. Hence, the perturbative approach is valid in the few-harmonic generation regime for low population of harmonic modes.

	The $\chi_n$ susceptibilities are not directly measurable, therefore we fit their values based on physical considerations. 
	The measured energy spectrum of the harmonics $\{I_n(|\alpha_0|,\tau)\}=\{\hbar\omega_n\langle N_n (|\alpha_0|,\tau)\rangle\}$ induced by a driving pulse of amplitude $|\alpha_0|=\sqrt{\epsilon_0 V/2\hbar\omega}E_0$ 
	and temporal duration $\tau$, can be used to determine $\chi_n$.
	Assuming that the zeroth-order approximation [Eq.~($\ref{classical}$)] acceptably predicts the harmonic energy spectrum, one can write 
	\begin{equation}\label{Intensitiyfitting}
		I_n(|\alpha_0|,\tau)=|\alpha_{0}|^{2n} \chi^2_n(|\alpha_0|) \hbar\omega_n \tau^2 ~,
	\end{equation}
	where we explicitly denoted the potential amplitude-dependence of the susceptibilities.
	Then $\chi_n$ can be expressed through experimentally measurable spectrum as
	\begin{align}\label{susceptibilitiesequ}
		\chi_n(|\alpha_0|) = \sqrt{\dfrac{I_n(|\alpha_0|,\tau)}{|\alpha_0|^{2n}\hbar\omega_n \tau^2}} ~.
	\end{align}
	We note that for analysis, one can choose some parameters freely, utilizing the scaling relations rooted in Eq.~(\ref{Intensitiyfitting}). 
	First, it follows that one can always scale up interaction time $\tau$ with the transformation
	\begin{equation}
		(\chi_n ,\tau) \to 
		(\tfrac{1}{\xi}\chi_n , \xi \tau) .
	\end{equation}
	Similarly, for a given harmonic energy $I_n$ and fixed interaction time $\tau$, only the product $|\alpha_0|^{2n}\chi^2_n$ needs to be invariant. Hence, one can choose arbitrary coherent amplitude $\alpha_0$ with corresponding  susceptibilities $\chi_n$, according to the transformation
	\begin{equation}\label{scaling}
		(\alpha_0, \chi_n\tau) \to
		(\xi \alpha_0 ,\tfrac{1}{|\xi|^n} \chi_n\tau ) .
	\end{equation}
	We note that the $\tau$ parameter can be eliminated by considering $\chi_n/\chi_m$ ratios, since
	\begin{align}
		\dfrac{\chi_n(|\alpha_0|)}{\chi_m(|\alpha_0|)} = \sqrt{\dfrac{I_n(|\alpha_0|,\tau)m}{I_m(|\alpha_0|,\tau)n}} \dfrac{|\alpha_0|^{m}}{|\alpha_0|^{n}}.
	\end{align}

	In realizations of HHG, odd-order harmonics often dominate the spectrum, that is, $\chi_n\neq0$ only when $n$ is an odd natural number. We stress that the "only-odd-harmonics" rule is strictly only valid for material targets with inversion symmetry, together with monochromatic excitation, for example atomic gases driven by quasi-monochromatic pulses \cite{MWENO05}. We will accept this assumption as valid in the following, but we emphasize that it can be abandoned without loss of generality.
	
	There are two aspects of the harmonic spectrum that are commonly used to categorize qualitative features of HHG: how the decrease of the harmonic energies depends on the increasing harmonic order, and the dependence of individual harmonic energies on the driving pulse amplitude.
	The literature of HHG distinguishes between the so-called perturbative regime and the nonperturbative regime. The idea  behind the concept of perturbative regime is that the harmonic energies $\{I_n\}$ in this regime scale with the laser pulse energy $I_0$ as $\propto I_0^n$ for low pump intensities, which behaviour is characteristic of perturbative processes. On the contrary, the spectrum in the nonperturbative regime deviates from the scaling characterizing the perturbative regime, and for sufficiently high intensity driving, generally characterized by a plateau-like harmonic spectrum.
		Nevertheless, the perturbative approach, as a calculational technique, can be used even in the nonperturbative regime, provided that only a few harmonics are generated and the harmonic mode populations remain low.
	Subsequently, we show how our model can be applied in both regimes. Then we examine how the model can be fitted to the experimental data measured in Ref.~\cite{PRXQuantum.5.040319}. Another example of parameter-fitting for the same model has been discussed in the Appendix of Ref.~\cite{PhysRevA.109.053717}.
	
	\subsection{Perturbative regime}
	
	The perturbative regime, physically corresponding to low driving intensities, is characterized by the domination of the lower-order harmonics. There are two defining features: First, harmonic energies fall off exponentially with increasing order of harmonics. We note that at very high orders the decrease can even be inverse factorial \cite{10.1063/1.339157}. The exponential decrease can be expressed as $I_n\propto C p^{2n}$, where $p\in[0,1)$ and $C\in\mathbf{R}^{+}$. Second, the harmonic energies scale with the driving amplitude as $I_n\propto|\alpha_0|^{2n}$.
	These two conditions together lead to
	\begin{equation}
		I_n(|\alpha_0|,\tau)= \tfrac{\hbar\omega_n}{n} C p^{2n} |\alpha_0|^{2n} \tau^2 , 
	\end{equation}
	from which, using Eq.~(\ref{susceptibilitiesequ}), we can express susceptibilities as
	\begin{equation}
		\chi_n(|\alpha_0|)=\sqrt{\dfrac{C}{n}} p^n.
	\end{equation}
	Accordingly, the properly determined susceptibilities in our model are independent of the driving pulse energy in the perturbative regime. 
	\begin{figure}[h!]
		\centering
		\includegraphics[width=1.0\linewidth]{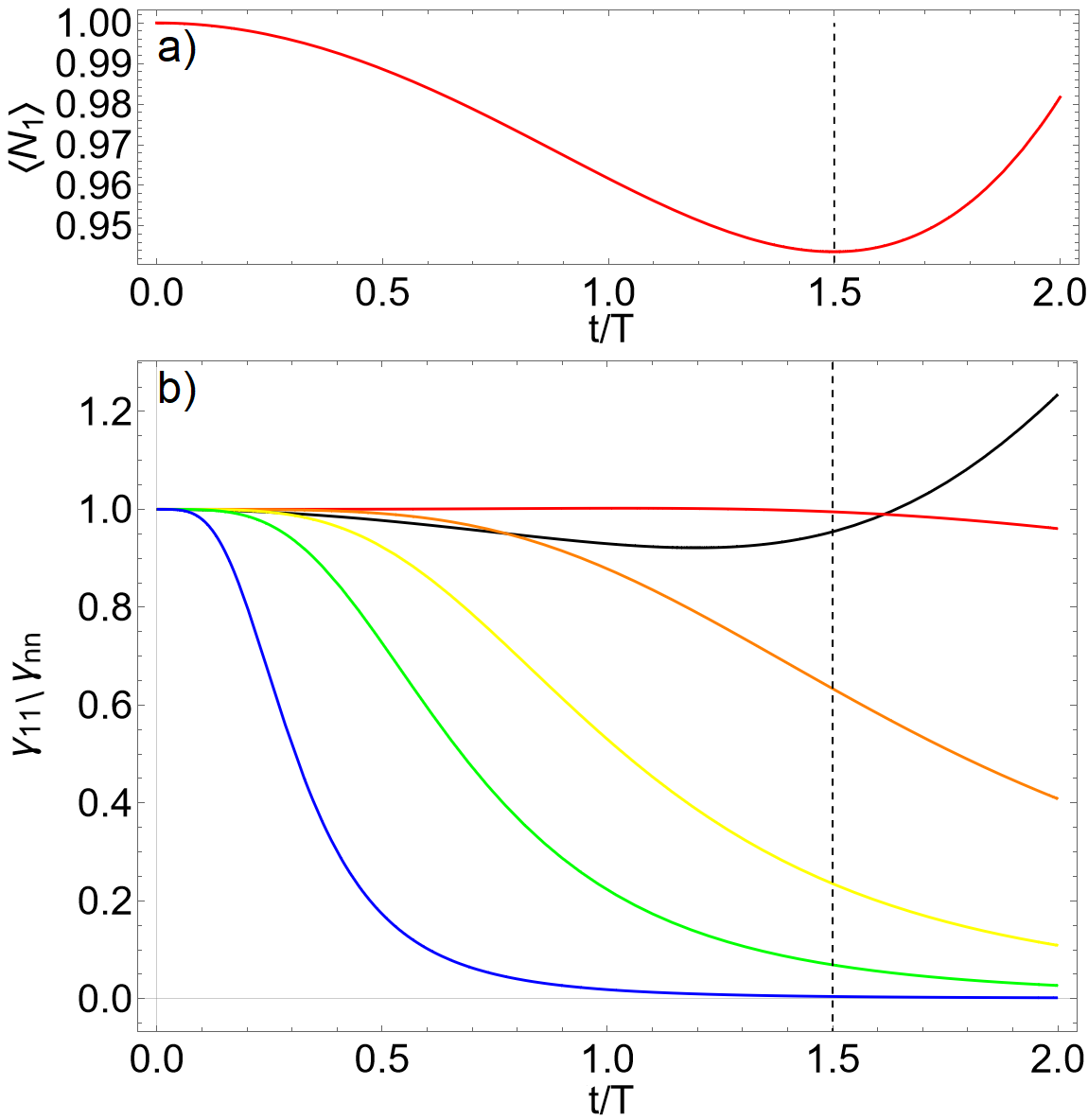}
		\caption{Single-mode physical quantities characterizing the excitation and harmonics. a) Time-evolution of $\langle N_1(t)\rangle$ shown as a function of $t/T$, where $T=2\pi/\omega$ is the optical cycle.
			b) Time-evolution of the photon correlations: $\gamma_{1,1}(t)$ (black), $\gamma_{3,3}$ (red); $\gamma_{5,5}(t)$ (orange); $\gamma_{7,7}(t)$ (yellow); $\gamma_{9,9}(t)$ (green); $\gamma_{11,11}(t)$ (blue). The parameters are chosen such that $|\alpha_0|=1$, $\chi_3=0.02$, $p=0.3$, with a cutoff order of $M=11$. Dashed lines show the chosen interaction time $t=1.5T$, applied in Fig.~(\ref{fig:vazlat3b})}
		\label{fig:vazlat3}
	\end{figure}
	
	Figure~\ref{fig:vazlat3} shows the physical quantities characterizing the excitation mode and harmonic modes, introduced in the second half of the previous section.
	The parameters are chosen such that the amplitude is $|\alpha_0|=1$, and $\chi_3=0.02$, $p=0.3$, with a cutoff order of $M=11$.
	Fig.~\ref{fig:vazlat3}(a) shows the time-evolution of mean photon value of the excitation mode $\langle N_1(t)\rangle$ given in Eq.~(\ref{drivingphotonnum}) as a function of dimensionless time $t/T$, where $T=2\pi/\omega$ is the optical cycle of the driving field. 
	The vertical dashed line shows the chosen interaction time of three-half optical cycles $t=1.5T$, which will be applied to calculate two-mode physical quantities. We note that such a choice practically corresponds to a monochromatic driving pulse with a rectangular carrier envelope. The interaction between the excitation and harmonic modes leads to an initial decrease of $\langle N_1(t)\rangle$, i.e. photon absorption from the driving mode, which is both intuitive and measurable. On the other hand, the subsequent increase corresponds to an energy transfer from the harmonics into the driving field, which practically cannot be measured experimentally. This phenomenon can pose a limitation for the physical validity of our model.
	\begin{figure}[h!]
		\centering
		\includegraphics[width=0.95\linewidth]{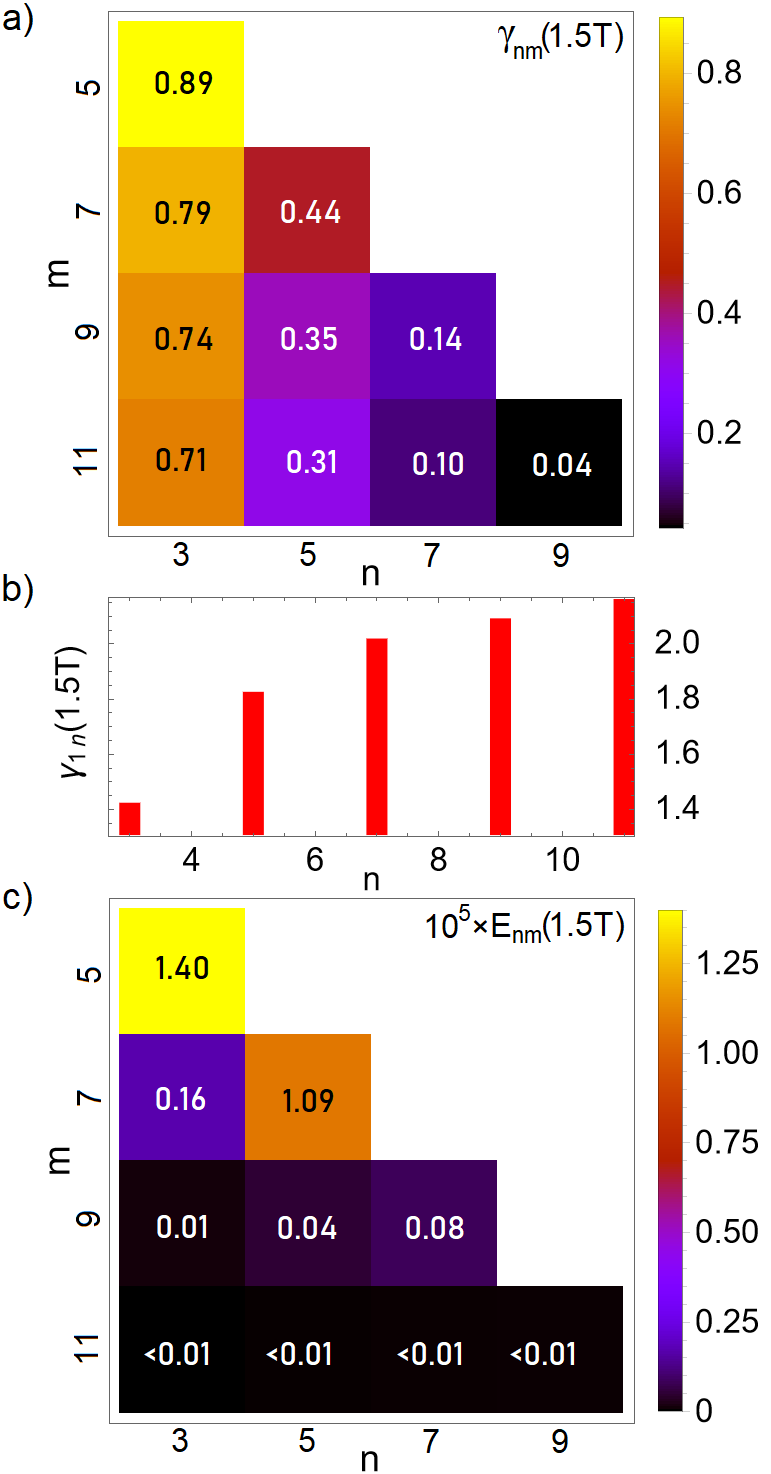}
		\caption{Two-mode physical quantities. Intermodal correlation function values $\gamma_{nm}(1.5T)$ and $\gamma_{1n}(1.5T)$ a) between harmonic modes and b) between the excitation mode and harmonic modes, respectively.
			c) Upscaled logarithmic negativity between harmonic modes $10^5\times E_{nm}(1.5T)$.}
		\label{fig:vazlat3b}
	\end{figure}
	
	The time-evolution of the photon autocorrelations $\gamma_{j,j}$ for the modes $j=1,3,\dots 11$ is presented in Fig.~\ref{fig:vazlat3}(b). 
	Comparing the values of correlation functions for the harmonics, they show a clear tendency of decreasing with both harmonic order and time. Harmonic modes are characterized by sub-Poissonian photon statistics, and this characteristic is especially significant for higher order harmonics. The correlation function of the pump mode $\gamma_{11}(t)$ has a moderately decreasing tendency with time, but close to the chosen interaction time $t=1.5T$, it starts increasing, representing a super-Poissonian statistics.
	
	The two-mode physical quantities are shown on Fig.~\ref{fig:vazlat3b}, calculated with identical parameters as were used for the single-mode quantities.
	Inter-harmonic correlation values at the end of the interaction $\gamma_{nm}(1.5T)$ are plotted in Fig.~\ref{fig:vazlat3b}(a). These are generally of lower than unit value, and the qualitative tendency is that correlation is larger if at least one harmonic is of lower order.
	Fig.~\ref{fig:vazlat3b}(b) presents the intermodal correlation values $\gamma_{1n}(1.5T)$ between the excitation mode and harmonic modes. Cross-correlations between the driving and harmonic modes are positive, which is what one expects for driving and driven modes. It can reach values $\gamma_{1n}>2$, primarily when the harmonics are of higher order. For high-order harmonic pairs, significant sub-Poissonian photon statistic, that is $\gamma_{nn}<1$, and simultaneous anti-correlations between these modes, characterized by $\gamma_{nm}<1$, are present, implying nonclassical quantum properties.
	To analyze pairwise entanglement, the logarithmic negativity between harmonics, expressed through Eq.~(\ref{lognegativit}), are shown on Fig.~\ref{fig:vazlat3b}(e), upscaled by a factor of $10^5$. The values quickly decrease as harmonic orders increase.
	Comparing Fig.~\ref{fig:vazlat3b}(a) and Fig.(\ref{fig:vazlat3b}(c), one can conclude, that lower-order harmonic pairs have more pronounced entanglement, while higher-order harmonic pairs are associated with more pronounced anti-correlations.
	
	Therefore, nonclassical correlations characterize both lower-order and higher-order pairs of harmonics, however, the exact nature of these nonclassicalities differ. 
	We note here that the calculated value $R_{3,5}(1.5T)\approx 1.2$ (not shown) can fall within the range of measured values for HHG using semiconductor materials in Ref.~\cite{PRXQuantum.5.040319}.
	
	\subsection{Idealized nonperturbative regime}
	The term nonperturbative regime --in the context of HHG-- typically refers to setups in which the driving intensity is of higher value. The harmonic energies in the resulting spectrum, after an exponential decrease at the lowest harmonics, reach a plateau-like, nearly constant value. This plateau extends over multiple harmonic orders before reaching a cutoff frequency. The exact structure of the plateau, which can be multi-leveled, varies depending on the material system. Due to the complexities involved, we will consider an idealized version of the nonperturbative regime, in which the harmonic energies are independent of the harmonic order. 
	This leads to:
	$I_n(|\alpha_0|,\tau)= \tfrac{\hbar\omega_n}{n} C  \tau^2$ and \begin{equation}
		\chi_n(|\alpha_0|) = \sqrt{\dfrac{C}{n}}\dfrac{1}{|\alpha_0|^{n}} ~,
	\end{equation}
	respectively. Unsurprisingly, the properly determined susceptibilities in our model are  intensity-dependent in the nonperturbative regime. 
	
\begin{figure}[h!]
		\centering	\includegraphics[width=0.97\linewidth]{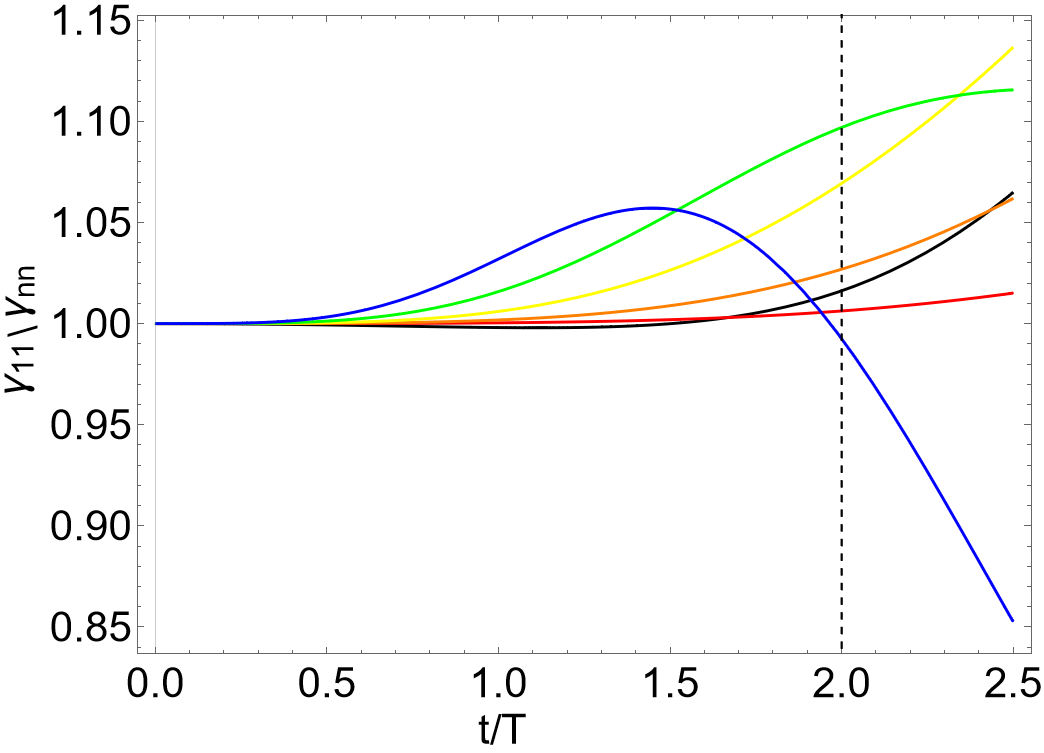}
		\caption{Time-evolution of the photon correlation functions, shown as a function of $t/T$, where $T=2\pi/\omega$ is the optical cycle. $\gamma_{1,1}(t)$ is shown as black, $\gamma_{3,3}$ red; $\gamma_{5,5}(t)$ orange; $\gamma_{7,7}(t)$ yellow; $\gamma_{9,9}(t)$ green; $\gamma_{11,11}(t)$ blue curves. Dashed lines show the corresponding interaction time $t=2T$. Parameters are $|\alpha_0|=\sqrt{20}$, $\chi_3=0.0002$, $M=11$. 
		}
		\label{fig:vazlat4}
	\end{figure}

	Fig.~\ref{fig:vazlat4} shows the time-evolution of the photon autocorrelation functions of the excitation mode $\gamma_{11}(t)$ as black; and of the harmonics $\gamma_{nn}$ as colored curves.
	The parameters were chosen to be $|\alpha_0|=\sqrt{20}$, and $\chi_3=0.0002$, with a cutoff order of $M=11$.
	The vertical dashed line shows the chosen interaction time of two optical cycles $t=2T$, which will be applied to calculate two-mode physical quantities. We note that such a choice practically corresponds to a monochromatic driving pulse with a rectangular carrier envelope.
	Comparing the values of correlation functions of the harmonics $\gamma_{nn}(t)$, they exhibit slightly super-Poissonian photon statistics, with a clear tendency of initial increase with both time and harmonic order. This dynamics markedly differs from that found in the perturbative regime. The initial increase is followed by a subsequent decrease, which takes place sooner for higher-order harmonics. The correlation function of the excitation mode $\gamma_{11}(t)$ have a moderately increasing tendency with time, representing a slightly super-Poissonian statistics at the chosen interaction time $t=2T$.
	\begin{figure}[h!]
		\centering
		\includegraphics[width=0.95\linewidth]{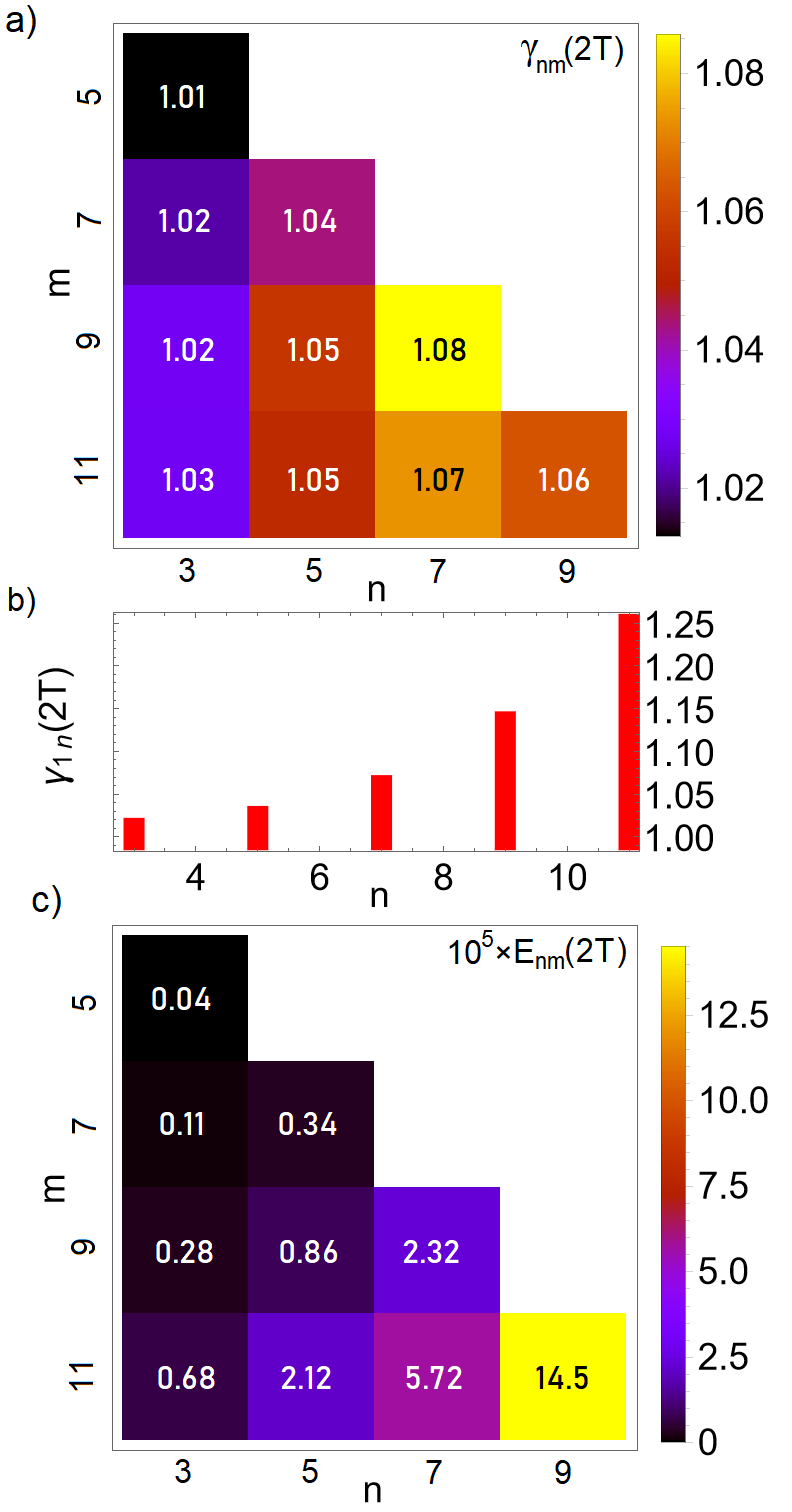}
		\caption{Two-mode physical quantities. Intermodal correlation function values $\gamma_{nm}(2T)$ and $\gamma_{1n}(2T)$ a) between harmonic modes and b) between the excitation mode and harmonic modes, respectively. c) Upscaled Logarithmic negativity between harmonic modes $10^5\times E_{nm}(2T)$.
		}
		\label{fig:vazlat4b}
	\end{figure}
	
	The two-mode physical quantities are displayed on Fig.~\ref{fig:vazlat4b}, calculated with identical parameters as used for the single-mode correlations.
	Inter-harmonic correlation functions $\gamma_{nm}(t)$, evaluated at interaction time $t=2T$, are plotted in Fig.~\ref{fig:vazlat3b}(a). These are generally of slightly above unit value, implying a moderately increased correlation between photon emissions into different harmonic modes. The qualitative tendency is that inter-harmonic correlation is more pronounced between higher-order harmonic pairs.
	Fig.~\ref{fig:vazlat3b}(b) presents the intermodal correlation values between the excitation mode and harmonic modes $\gamma_{1n}(2T)$. These values are positive, but lower than in the perturbative regime. Comparing these two figures, it can be observed that all calculated cross-correlations, both excitation-harmonic and harmonic-harmonic ones have lower values than 2.

	To analyze pairwise entanglement, the logarithmic negativity between harmonics, expressed through Eq.~(\ref{lognegativit}) have been plotted on Fig.~\ref{fig:vazlat3b}(c), scaled up by a factor of $10^5$. One can see that the inter-harmonic entanglement are more pronounced between higher-order harmonics.
	It can be observed, that  entanglement is more pronounced within high-order harmonic pairs.
	While we will not delve into potential tripartite (or more general multipartite) entanglement in this study, we would like to point out that it is possible to find three harmonic modes (a, b, c) within the spectrum that simultaneously exhibit $E_{a,b}>0$, $E_{a,c}>0$, and $E_{b,c}>0$. This indicates that, at least, entanglement between bipartite reductions is present.
	
	In the perturbative regime, the logarithmic negativity (Fig.3/c) is highest between the lowest-order harmonics. This indicates that entanglement is strongest among low-order harmonics, which also dominate the spectrum as mean photon numbers decay exponentially with the harmonic order.
	Conversely, in the idealized non-perturbative regime, the trend of the logarithmic negativities (Fig.5/c) reverses: the logarithmic negativity is highest between the highest-order harmonics. 
	This coincides with the mean photon numbers showing a moderate increase with the harmonic order.
	The general conclusion is that for all investigated regimes, the degree of entanglement is positively correlated with mode population, i.e. higher average photon numbers indicate a higher degree of entanglement between the respective modes.
	
	\subsection{Fitting to experimental results}
	Now we turn our attention to an analysis of the experimental measurements reported in Ref.~\cite{PRXQuantum.5.040319}. In this experiment, a pulse of intermediate intensity was used, and the $3$rd and $5$th order harmonics were studied for GaAs, ZnO and Si targets. Photons-per-pulse associated with these modes, as well as their auto-; and cross-correlation functions $\gamma_{33}$, $\gamma_{55}$ and $\gamma_{35}$, respectively, have been measured, and $R_{35}$ have been determined. These quantities were examined as functions of the driving intensity $\hbar\omega\tfrac{|\alpha_0|^2 c}{V}$, where $V$ is the interaction volume. The driving intensity, of course, is proportional to the pulse energy $\hbar\omega|\alpha_0|^2$.
	The process of harmonic generation within the used range of laser intensities contains both the perturbative regime, and the onset of a low-intensity nonperturbative regime. 
	The transition between these two regimes occur relatively suddenly, around a certain $|\alpha_0|=\alpha_{(n)}$ driving amplitude, which depends on the harmonic order $n$. The measured harmonic energies scale with the driving amplitude as $I_n\propto|\alpha_0|^{\epsilon_n}$, with exponent $\epsilon_n<2n$ in the nonperturbative regime. 
	
	We fit the susceptibilities through the measured mean photons per pulse in the harmonic modes, using the zeroth order solution Eq.~(\ref{classical}), as in the previous subsections, and assume that the laser pulse energy can be written as $I_0=\hbar\omega |\alpha_0|^2$. Similarly, we express the mean photons per pulse associated with the $n$th harmonic as $I_n/\hbar\omega_n$. These assumptions are justified by the fact that the measured mean photons are low, corresponding to a low $\chi_n t$ value in Eq.~(\ref{meanphotonnharm}). Hence, the leading-order correction in the mean photon-number due to deviation from a coherent state is small. Thus, one can safely approximate the photons per pulse measurements $\tfrac{I_n}{\hbar \omega} \approx \chi_n^2 \tau^2 |\alpha_0|^{2}$ to fit the $\chi_n$ susceptibilities. Given the scaling $I_n\propto|\alpha_0|^{\epsilon_n}$, we can proceed as
	\begin{align}\label{derivefitting}
		\chi_n^2(|\alpha_0|) \tau^2 |\alpha_0|^{2n} =
		C_n \tau^2 |\alpha_0|^{\epsilon_n} ~,
		\\ 
		\label{fittinguse} 
		\langle N_n(\tau)\rangle \approx 
		C_n \tau^2 \left(\dfrac{I_0}{\hbar\omega}\right)^{\epsilon_n/2}.
	\end{align}
	From Eq.~(\ref{derivefitting}) we can derive the susceptibilities as $\chi_n(|\alpha_0|)=\sqrt{C_n/|\alpha_0|^{2n-\epsilon_n}}$. These susceptibilities are, of course, intensity-dependent in the nonperturbative regime, in agreement with the previous subsection. Equation (\ref{fittinguse}) can be used to determine $\epsilon_n$ values from the photon-per-pulse measurements.
	Using data from the figures Fig.~1(c), Fig.~S3(b), and Fig.~S4(b) of Ref.~\cite{PRXQuantum.5.040319}, one can obtain that $\epsilon_3 \approx 4.6$, and $\epsilon_5\approx6$ for GaAs target, $\epsilon_3 \approx 4$ and $\epsilon_5\approx6$ for ZnO target, and $\epsilon_3 \approx 5.6$ and $\epsilon_5\approx5.8$ for Si target, respectively.
	The ratio $C_3/C_5$ can also be estimated from these figures.
	To proceed, we express the effective susceptibilities in a way that incorporates the transition between perturbative and low-intensity nonperturbative regimes. For this purpose, we write
	\begin{equation}
		\chi_n(|\alpha_0|)= 
		\begin{Bmatrix}
			\chi_n^{(pert)} & \text{if} ~~|\alpha_0|\leq \alpha_{(n)}
			\\
			\chi_n^{(pert)} \left(\frac{\alpha_{(n)}}{|\alpha_0|}\right)^{n-\tfrac{\epsilon_n}{2}} & \text{if} ~~|\alpha_0|>\alpha_{(n)}
		\end{Bmatrix}.
	\end{equation}
	With the above considerations, susceptibilities are chosen so that they replicate the $\langle N_n(\tau) \rangle$ dependencies on the pulse energy reasonably well, while keeping in mind the limits of validity of the perturbative calculations.

\begin{figure}[!]
		\centering
		\includegraphics[width=1.0\linewidth]{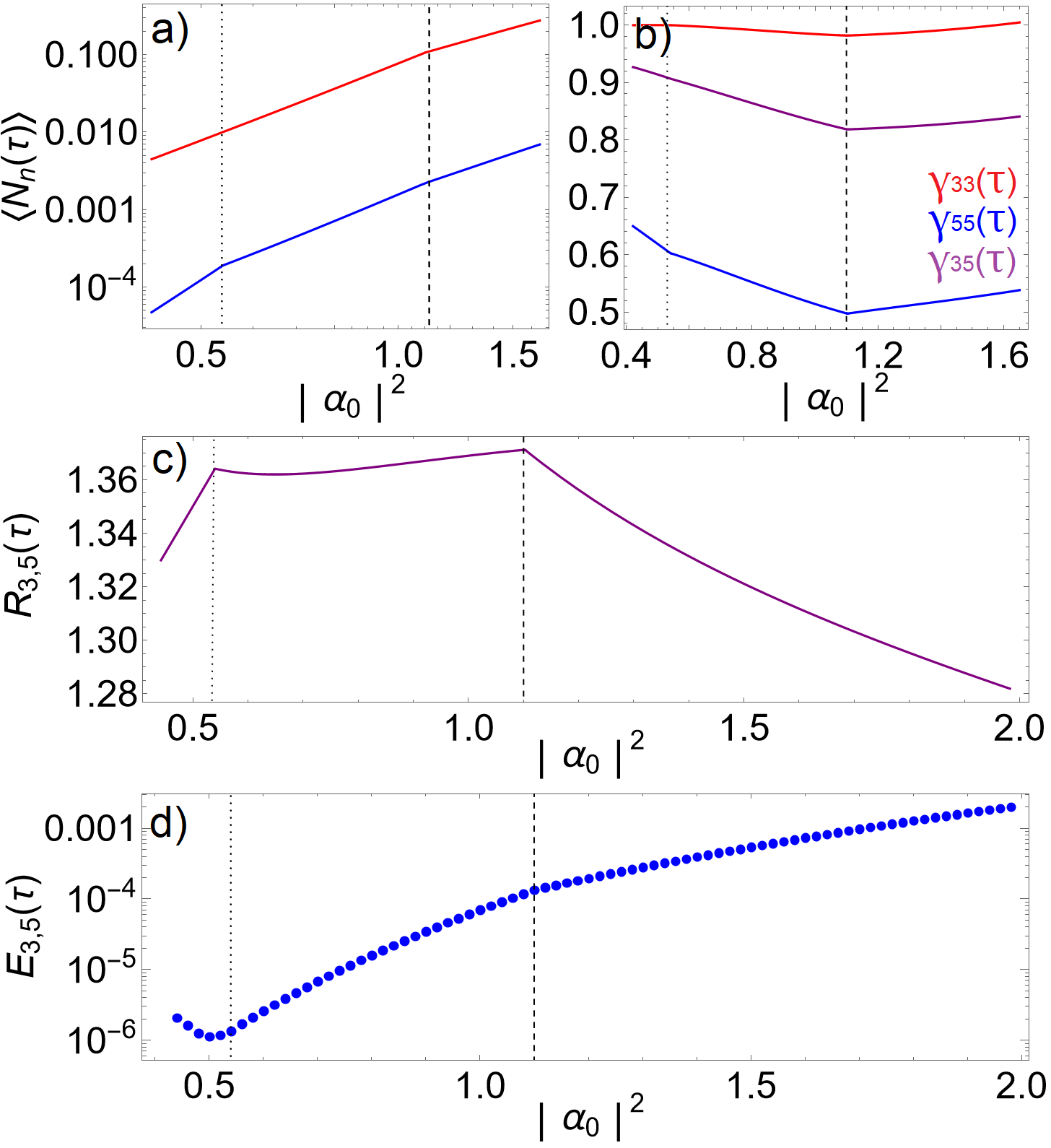}
		\caption{Physical quantities as function of $|\alpha_0|^2$, using GaAs susceptibilities given by equations (\ref{GaAs}-\ref{GaAs2}). Dashed and dotted lines show the transitions to the non-perturbative regime for the 3'rd and 5'th harmonics, respectively. a) Mean photon numbers $\langle N_3(\tau)\rangle$ (red) and $\langle N_5(\tau)\rangle$ (blue). b) Shows the photon auto-; and cross-correlation functions at the end of the interaction. 
			Subfigure c) shows the $R_{3,5}(\tau)$ values. d) The logarithmic negativity $E_{3,5}(\tau)$.
		}
		\label{fig:vazlat5gaas}
	\end{figure}

\begin{figure}[h!]
		\centering
		\includegraphics[width=1.0\linewidth]{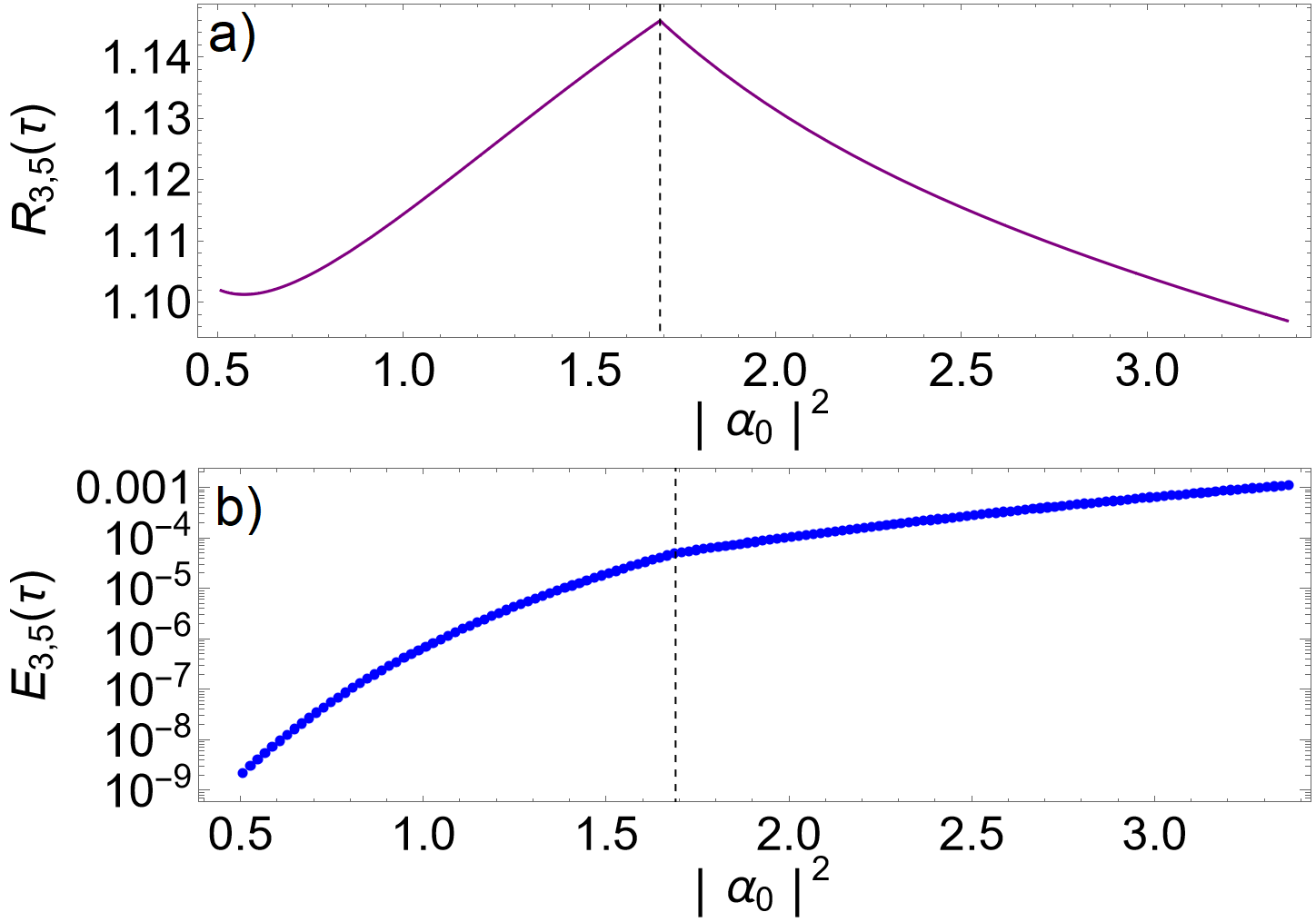}
		\caption{Physical quantities as function of $|\alpha_0|^2$, using ZnO susceptibilities given by equations (\ref{ZnO}-\ref{ZnO2}). a) $R_{3,5}(\tau)$ value and b) logarithmic negativity $E_{3, }(\tau)$, respectively.}
		\label{fig:vazlat5zno}
	\end{figure}

\begin{figure}[h!]
		\centering
		\includegraphics[width=1.0\linewidth]{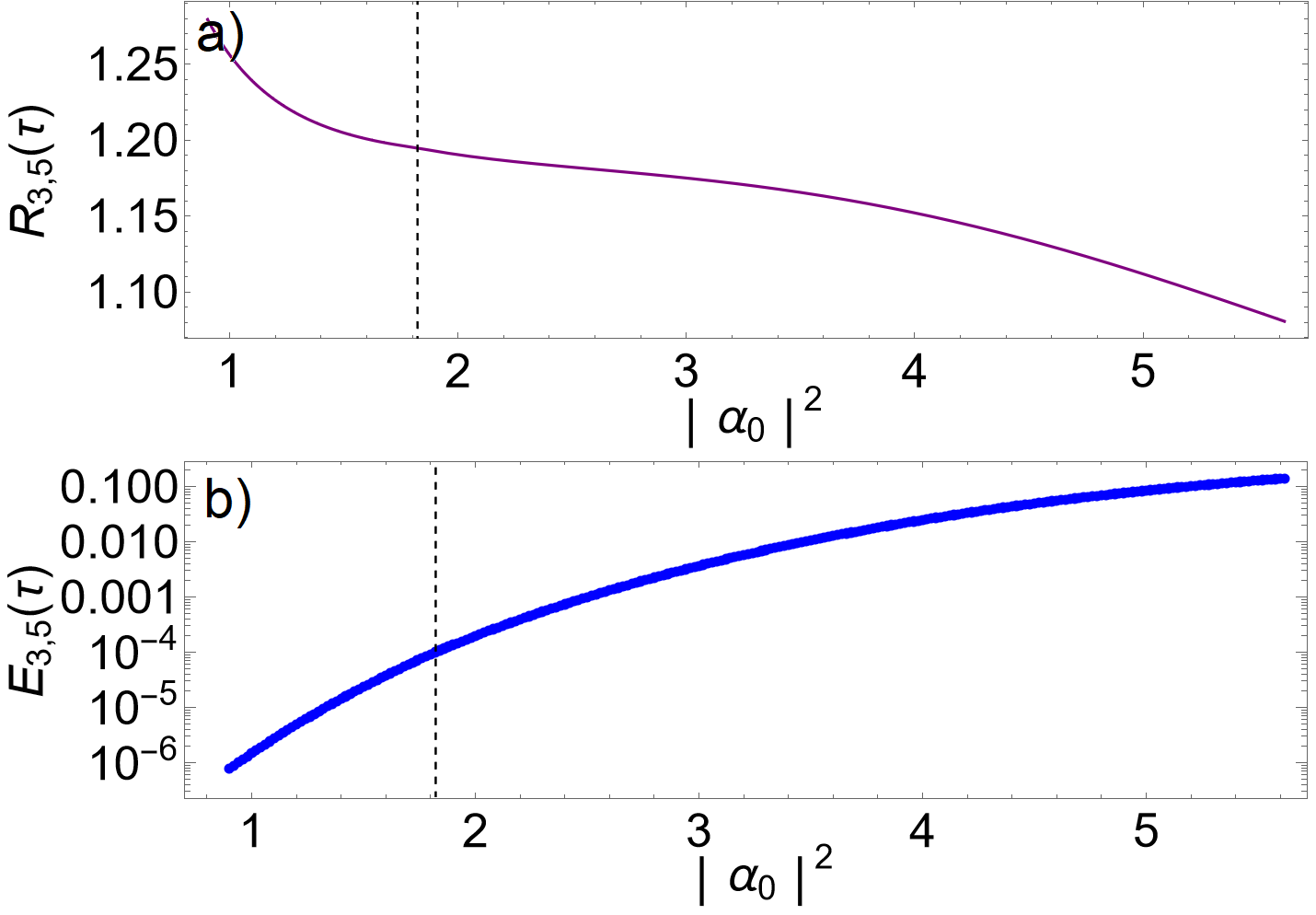}
		\caption{Physical quantities as function of $|\alpha_0|^2$, using Si susceptibilities given by equations (\ref{Si}-\ref{Si2}). The same quantities are listed as in Fig.~\ref{fig:vazlat5zno}. }
		\label{fig:vazlat5si}
	\end{figure}

	In the following, we will assume that the driving pulse is of half-cycle duration, that is, $\tau=T/2$. 
	Recall that while we have chosen the range of $|\alpha_0|$ to be of small values, it can be scaled up arbitrarily. 
	The susceptibilities we have chosen for the three materials are listed in equations~(\ref{GaAs}-\ref{Si2}). Using these and the formulae of Sec.~\ref{analyticalresults}, we have calculated the physical quantities measured in Ref.~\cite{PRXQuantum.5.040319}. The results of our analysis are shown on Fig.~(\ref{fig:vazlat5gaas}), Fig.~(\ref{fig:vazlat5zno}) and Fig.~(\ref{fig:vazlat5si}), showing physical quantities as functions of $|\alpha_0|^2$, which read
	\begin{align}\label{GaAs}
		\chi^{\text{(GaAs)}}_3(|\alpha_0|)= 
		\begin{Bmatrix}
			0.069 & \text{if} ~~|\alpha_0|\leq 1.05
			\\
			0.069\left( \frac{1.05}{|\alpha_0|}\right)^{0.7} & \text{if} ~~|\alpha_0|> 1.05
		\end{Bmatrix},
		\\ \label{GaAs2}
		\chi^{\text{(GaAs)}}_5(|\alpha_0|)=
		\begin{Bmatrix}
			0.01 & \text{if} ~~|\alpha_0|\leq 0.735
			\\
			0.01 \left( \frac{0.735}{|\alpha_0|}\right)^{2} & \text{if} ~~|\alpha_0|> 0.735
		\end{Bmatrix},
	\end{align} \vspace{-0.9cm}
	\begin{align}\label{ZnO}
		\chi^{\text{(Zno)}}_3(|\alpha_0|)= 
		\begin{Bmatrix}
			0.041 & \text{if} ~~|\alpha_0|\leq 1.3
			\\
			0.041\left( \frac{1.3}{|\alpha_0|}\right)^{1} & \text{if} ~~|\alpha_0|> 1.3
		\end{Bmatrix},
		\\ \label{ZnO2}
		\chi^{\text{(Zno)}}_5(|\alpha_0|)=
		\begin{Bmatrix}
			0.035 & \text{if} ~~|\alpha_0|\leq 0.377
			\\
			0.035 \left( \frac{0.377}{|\alpha_0|}\right)^{2} & \text{if} ~~|\alpha_0|> 0.377
		\end{Bmatrix},
		\\ \label{Si}
		\chi^{\text{(Si)}}_3(|\alpha_0|)= 
		\begin{Bmatrix}
			0.027 & \text{if} ~~|\alpha_0|\leq 1.35
			\\
			0.027\left( \frac{1.35}{|\alpha_0|}\right)^{0.2} & \text{if} ~~|\alpha_0|> 1.35
		\end{Bmatrix},
		\\ \label{Si2}
		\chi^{\text{(Si)}}_5(|\alpha_0|)=
		\begin{Bmatrix}
			4.31 & \text{if} ~~|\alpha_0|\leq 0.06
			\\
			4.31 \left( \frac{0.06}{|\alpha_0|}\right)^{2.1} & \text{if} ~~|\alpha_0|> 0.06
		\end{Bmatrix}.
	\end{align}
	
	Comparing Fig.~\ref{fig:vazlat5gaas}(a) to Fig.~1(b) of Ref.~\cite{PRXQuantum.5.040319}, one can see that the measured mean photons-per-pulse qualitatively corresponds to the calculated mean photon numbers $\langle N_3(\tau)\rangle$ and $\langle N_5(\tau)\rangle$, shown as red and blue curves on Fig.~\ref{fig:vazlat5gaas}(a). The parameters have been chosen so that the ratio $\langle N_3(\tau)\rangle/\langle N_5(\tau)\rangle$ and the ratio of transition amplitudes $\alpha_{(3)}/\alpha_{(5)}$ correspond to the measured values.
	Fig.~\ref{fig:vazlat5gaas}(b) shows the values of the auto-; and cross-correlation functions $\gamma_{33}(\tau)$, $\gamma_{55}(\tau)$ and $\gamma_{35}(\tau)$, respectively. The generally decreasing tendency with the pulse energy is reproduced up to the transition amplitude of the $3$rd order harmonic $\alpha_{(3)}$.   
	However, the values of these correlation functions are typically below 1, unlike the measured values. 
	Nevertheless, valuable insight can be gained from the intensity-dependence of the correlation function: At transition amplitude $\alpha_{(n)}$, not only $\gamma_{nn}$, but the autocorrelation function of the other mode also experiences a transition. This can also be observed to a lesser degree on plots of mean photon numbers $\langle N_n(\tau)\rangle$. The explanation is straightforward when observing Eq.~(\ref{meanphotonnharm}) and Eq.~(\ref{gnn}). Since $\Omega_{n,0}$ contains all $\chi_n$ susceptibilities [see Eq.~(\ref{omega}) and Eq.~(\ref{theta})], any transition in the largest susceptibilities will be imprinted in all photon numbers and related quantities. This observation may explain the simultaneous deviations of the measured photons-per-pulse of the $3$rd and $5$th harmonic, observable in Fig.~S3(b) of Ref.~\cite{PRXQuantum.5.040319}.
	
	We present the $R_{3,5}(\tau)$ values on Fig.~\ref{fig:vazlat5gaas}(c), Fig.~\ref{fig:vazlat5zno}(a), and Fig.~\ref{fig:vazlat5si}(a)
	The CBS-inequality violation, that is $R_{3,5}(\tau)>1$, is reproduced as measured, implying the presence of nonclassical correlations in the  pair of harmonic modes. One can see that the material-dependent $\alpha_{(n)}$ transition amplitudes can have a pronounced effect on the intensity-dependence of $R_{3,5}(\tau)$. While the calculated values are somewhat larger than the measured ones for the chosen parameters, the qualitative agreement is easy to check when comparing Fig.~\ref{fig:vazlat5gaas}(c) to Fig.~3 of Ref.~\cite{PRXQuantum.5.040319}, and analogously Fig.~\ref{fig:vazlat5zno}(a) and Fig.~\ref{fig:vazlat5si}(a) to Fig.~S6 of Ref.~\cite{PRXQuantum.5.040319}.
	Somewhat surprisingly, a plateau-like structure can be reproduced for the case of GaAs [Fig.~\ref{fig:vazlat5gaas}(c)] between the transition pulse energies $|\alpha_{(5)}|^2$ and $|\alpha_{(3)}|^2$. 
	For the case of ZnO, our model can imitate the moderately increasing tendency of $R_{3,5}(\tau)$ up to $|\alpha_{(3)}|^2$ pulse energy, followed by a slow decline [Fig.~\ref{fig:vazlat5zno}(c)]. For the Si target, our model predicts monotonously decreasing $R_{3,5}(\tau)$ values with increasing pulse energy, without particularly notable transition.
	
	While nonclassical correlations have been both measured and can be reproduced by our calculation, it is not straightforward that the nonclassicality implies entanglement. We present the logarithmic negativities $E_{3,5}(\tau)$ as functions of pulse energy on Fig.~\ref{fig:vazlat5gaas}(d), Fig.~\ref{fig:vazlat5zno}(b), and Fig.~\ref{fig:vazlat5si}(b). 
	The logarithmic negativity generally has an increasing tendency with the pulse energy, although for low pulse energies we observed decreasing tendencies as well. Generally, the $R_{3,5}(\tau)$ values and the $E_{3,5}(\tau)$ functions have little correlation with increasing. However, in cases of "well-structured" $R_{3,5}(\tau)$ functions [such as in Fig.~\ref{fig:vazlat5gaas}(d)] the transition pulse energies are, to a small degree, imprinted on the negativity function.
	We note that the logarithmic negativities have not been measured directly, but our calculations support the interpretation that the arising nonclassicalities are, at least partially, in the form of entanglement.
	
	Finally, we note that the problem of fitting parameters to perturbative results often becomes a trade-off between the validity of the results and the fidelity with which the data is reproduced.
	Furthermore, let us add that in some experiments, such as Ref.~\cite{Rasputnyi2024}, the mean photon number of the driving pulse has been measured, instead of its intensity, leading to slightly different, although similar expressions.
	
	\section{Discussion and outlook}\label{discussoutlook}
	
	First, the physical applicability and limitations of the effective model merits some discussion.
	While describing a medium via effective susceptibilities is a well-established approach in quantum optics, it is not self-evident that the key features of HHG can be captured this way, given that the HHG process is fundamentally driven by strongly nonlinear electron dynamics.
	To precisely delineate the domain of validity for the effective model, a rigorous derivation of the Hamiltonian given in Eq.~({\ref{parHam}}) from a first-principles approach -wherein the driving field, harmonics and target material are simultaneously quantized, see e.g., Ref.~\cite{photonics8070269}- is required. 
	Once that is done, the domain of validity can be defined by the set of underlying approximations used during the derivation. 
	To establish a formal link between microscopic parameters and effective susceptibilities, one must start from a microscopic model, proceeding to take expectation values with respect to microscopic wavefunctions, and subsequently perform spatial averaging over the interaction volume. While this is beyond the scope of the current work, we offer a reasonable heuristic argument:
	Assuming that the interaction leaves the target material (i.e., electrons) in a final state close to its initial state, tracing out the material degrees of freedom would certainly yield a Hamiltonian similar to ours, basically due to the conservation of energy.

	Our Hamiltonian is expected to be valid for low and intermediate driving intensities, that is, in the regime of few-harmonic generation. Traditionally, this lower intensity regime has been less explored within classical descriptions of HHG, although it represents a domain where nonclassical light sources are relevant for quantum information technologies. It is worth noting that recent experimental analysis of the quantum-optical aspects have also been performed in this regime.
	Assuming the material is treated in a quantized manner, the regime of validity corresponds to the previously mentioned assumption that the final and initial states are identical, together with a negligible ionization rate, and a small number of energy-levels contributing to the nonlinear optical response.
	
	It may be instructive to consider the applicability of our phenomenological model for treating well-known frameworks in the field of HHG, such as the strong-field approximation (SFA) used for gas-phase targets \cite{Symphony,43047dc146e64dd1980cdee6d66fe5b8,BMB05,PhysRevA.104.033117}, and the semiconductor Bloch equations used for solid-state systems  \cite{WGRK15,PhysRevA.109.012223,TCHM20,Foldi17}. 
	The SFA approach assumes that the ionization plays a key role, occuring in the regime of high driving intensities. In this approach, the electron in the continuum is treated as a driven free particle. Bloch-equation-based approaches assume that incorporating a finite number of energy bands into the calculation is sufficient to describe the interband and intraband dynamics of the driven electron.
	These features suggest that the conditions for the applicability of our phenomenological model can be satisfied within a weakly driven Bloch-equation approach, whereas these conditions are incompatible with the assumptions of the SFA. 
	On the other hand, standard predictions, such as the presence of a plateau and the intensity-dependence of the cutoff frequency, can be incorporated empirically into our model through an appropriate fitting  of the susceptibilities, as discussed above.
	
	
	The question of interpretation, regarding the physical origin of the emerging entanglement, naturally arises. The origin of the resulting correlations and entanglement in our model is basically due to the interaction between the harmonics and the driving field. The driving field, through the medium, is coupled to all harmonics modes. Consequently, one can consider all the harmonic modes to be coupled to each other indirectly, through the driving field mode. That such a dynamics could lead to correlations between the harmonics --coinciding with depletion from the driving field-- is unsurprising and  intuitively sound.
	A broad implication of our calculation is that the back-action-induced inter-harmonic entanglements and the non-Gaussian characteristics of the individual harmonic quantum states are intrinsically linked.
	Since this process occurs in a diverse range of systems, one can expect that HHG in general can be used for engineering inter-harmonic entanglements.



	From an experimental perspective, however, quantifying the entanglement rigorously remains a challenge. Experimentally extracting the logarithmic negativity requires density matrix reconstruction of two harmonic modes via joint quantum state tomography; see, e.g., Ref.~\cite{s41534-025-00995-1}. In the regime analyzed here, the logarithmic negativities are small compared to those of maximally entangled states, imposing high sensitivity on the detection. 
	Nevertheless, our model predicts that nonclassical intermodal correlations strengthen with increasing photon numbers, offering a promising direction for future experiments. 
	While joint quantum state tomography is experimentally involved, when realized, the retrieved Wigner function and calculated logarithmic negativity is directly comparable to our results. 
	Alternatively, using entanglement witnesses --inequalities based on the measurement of commuting observables-- can be experimentally more accessible. 
	If one merely wishes to establish the presence of entanglement between different harmonic modes, the Duan inseparability criterion can be checked experimentally in a straightforward setup \cite{nphoton.2013.340}.
	The most direct experimental access to the quantum features is through photon correlation measurements. A primary nonclassical signature is accessible via photon statistics measured in a Hanbury Brown and Twiss configuration, where the analyzed correlation functions are directly observable,
	Furthermore, Cauchy-Schwarz inequality violations based on these correlation functions can detect global or local nonclassicality in two-mode states, which under specific conditions confirms the presence of entanglement. 
	Correlation measurement is limited in practice only by the picosecond temporal resolution of the detectors. 
	Additionally, experimental challenges to be addressed include the minimization of phonon-related modifications by cooling the samples, and using  thin samples to avoid reabsorption and scattering effects.


	Finally, because the resulting quantum-state of the harmonics resembles a generalized displaced W-state, one can presume that potential applications would be similar to that of the W-state on photonic modes of different frequencies. 
	Entanglement distributed across distinct optical frequencies is attractive due to its compatibility with fiber-optic networks and its resilience against decoherence in noisy channels \cite{PhysRevA.111.042407}.
	By converting these multi-frequency photonic $W$ states into long-lived excitations via optomechanical interactions, the quantum states can be placed into delayed storage \cite{PhysRevA.111.042407}.
	Similar realizations enable frequency-multiplexed quantum memories, allowing simultaneous entanglement swapping operations within a single fiber channel via wavelength-division multiplexing \cite{PhysRevLett.113.053603}.

	\section{Conclusions}\label{conclusion}
	
	We have considered an effective quantum optical model of HHG, in which both the driving field and the harmonic radiation are quantized, while the material is included only through susceptibilities. Analogously to traditional models of harmonic generation, we have assumed that the dominant process is the simultaneous absorption of $n$ photons from the driving field and emission of a photon into the $n$th harmonic mode. 
	We have performed a perturbative calculation and analytically evaluated quantum optical quantities associated with single modes, and with pairs of modes. We have analyzed the behaviour of these quantities for various regimes of the HHG process.
	
	We have found that nonclassicalities, comparable in magnitude to experimental measurements, emerge. These nonclassical properties include entanglement between harmonic modes and can be present in the few-harmonic generation regime. Furthermore, we have found that by fitting susceptibilities to the data of a recent experiment \cite{PRXQuantum.5.040319}, our model could, for certain materials, reproduce characteristics of the intensity-dependence of the nonclassical correlation measure $R$.
	
	As our model neglects the quantum state of the target material, the interpretation of the results is straightforward: By incorporating the depletion of the driving field during HHG, the back-action on the pump is naturally incorporated into our model. Consequently, the back-action modifies the quantum state of the driving field, leading to a modification in the quantum state of the harmonic modes. Ultimately, as a result of this, intermodal correlations and entanglement between the harmonics emerge.
	
	Our phenomenological approach has the advantage of not depending on specific properties of the target medium, except for the effective $\chi$ susceptibilities. Therefore, any feature of HHG that can be reproduced within our model can be considered a general aspect of HHG, not specific to the interaction material.
	We believe that theoretical research in this field can benefit from our results, and provide further insight into the quantum aspects of HHG.

	\section*{ACKNOWLEDGMENTS}
	We are thankful to Péter Földi and Sándor Varró for the useful suggestions with respect to the conceptual framework.
	This research was supported by the National Research, Development and Innovation Office, Hungary ("Quantum Information National Laboratory of Hungary" Grant No. 2022-2.1.1-NL-2022-00004, "Frontline" Research Excellence Programme Grant No. KKP133827, and Grant No. TKP2021-EGA-17).
	
	\appendix

	\section{Quantum optical quantities in perturbative approach}\label{quantities}
	
	Below, we present the derivation of the quantum optical quantities used in Sec. \ref{analyticalresults}, in a more general form. Here, we do not assume the initial condition used in the second half of section \ref{analyticalresults}, and we let the lowest-order harmonic symbolically be $n_1$ instead of 2. The square norm is straightforward to express
	$ \mathcal{N}(t) =
	1 + \sum_{k=1}^{M} |\Theta_k|^2 t^4
	+ \sum_{k=0}^{M} \sum_{j=n_1}^{M} |\Omega_{j,k}|^2 t^6$.
	Below, we first give the mean photon numbers of the driving mode
	\begin{align}
		\langle N_1 (t) \rangle = \dfrac{\langle \Psi| N_1 |\Psi \rangle}{\langle \Psi|\Psi \rangle} 
		\nonumber \\
		= \dfrac{1}{\mathcal{N}(t)} \langle \Psi'| N_1 + \alpha_{0}e^{-i\omega t}a^\dagger_1 + \alpha^*_{0}e^{i\omega t}a_1 + |\alpha_0|^2 |\Psi' \rangle
		\nonumber \\
		=  \dfrac{1}{\mathcal{N}(t)} 
		\bigg[ |\alpha_0|^2 \mathcal{N}(t) 
		- (\Theta_1 \alpha^*_0 + \Theta^*_1 \alpha_0 ) t^2 
		\nonumber \\
		+ \sum^{M}_{k=1} k|\Theta_k|^2 t^4 
		+ \sum^{M}_{k=0}\sum^{M}_{j=n_1}  k|\Omega_{j,k}|^2 t^6 
		\nonumber \\
		+  \sum_{k=1}^{M} \sqrt{k+1}(\alpha_{0} \Theta^*_{k+1}\Theta_{k} 
		+ \alpha^*_{0} \Theta^*_{k}\Theta_{k+1} ) t^4 ~+
		\nonumber 
	\end{align} \begin{align}
		\sum_{k=0}^{M} \sum_{j=n_1}^{M}
		\sqrt{k+1}(\alpha_{0} \Omega^*_{j,k+1}\Omega_{j,k} + \alpha^*_{0} \Omega^*_{j,k}\Omega_{j,k+1} ) t^6
		\bigg]	,
	\end{align}
	and the mean photon number of the $n'$th harmonic mode
	\begin{align}
		\langle N_n (t) \rangle = 
		\dfrac{1}{\mathcal{N}(t)} \langle \Psi| N_n |\Psi \rangle 
		\nonumber \\
		= \dfrac{1}{\mathcal{N}(t)}  \langle \Psi'| N_n 
		- it\chi_n \alpha^n_{0}e^{-in\omega t}a^\dagger_n 
		+ it\chi_n \alpha^{*n}_{0}e^{in\omega t}a_n 
		\nonumber \\
		+ t^2\chi^2_n|\alpha_{0}|^{2n} |\Psi' \rangle
		= \dfrac{1}{\mathcal{N}(t)} \bigg[ 
		\chi^2_n|\alpha_{0}|^{2n} t^2 \mathcal{N}(t) \nonumber\\
		+ \sum^{M}_{k=0} |\Omega_{n,k}|^2 t^6
		+ \chi_n (\alpha^n_{0} \Omega^*_{n,0}  
		+ \alpha^{*n}_{0} \Omega_{n,0})t^4 
		\nonumber\\
		- \chi_n \sum_{k=1}^{M} (\alpha^n_{0}\Theta_{k} \Omega^*_{n,k}
		\!+ \alpha^{*n}_{0} \Theta^*_{k} \Omega_{n,k}) t^6
		\bigg] ,
	\end{align}
	respectively. The coherence functions are listed below. Autocoherence of the driving field mode is
	\begin{align}
		G_{11}(t)=
		\dfrac{1}{\mathcal{N}(t)} \langle \Psi'| 		(a^\dagger_1 + \alpha^*_{0}e^{i\omega t} )
		(a^\dagger_1 + \alpha^*_{0}e^{i\omega t} )
		\nonumber \\
		(a_1 + \alpha_{0}e^{-i\omega t} )
		(a_1 + \alpha_{0}e^{-i\omega t} )  |\Psi' \rangle 
		\nonumber \\
		= \dfrac{1}{\mathcal{N}(t)} \langle \Psi'| 
		a^{\dagger2}_1 a^{2}_1 
		+ 2(\alpha e^{-i\omega t} a^{\dagger2}_1 a_1 + \alpha^* e^{i\omega t} a^{\dagger}_1 a^2_1 )
		\nonumber \\	
		+ \alpha^2_0 e^{-i2\omega t} a^{\dagger2}_1 + \alpha^{*2}_0 e^{i2\omega t} a^{2}_1
		+ 4 |\alpha_0|^2 a_1^\dagger a_1
		\nonumber \\ 
		+ 2|\alpha_0|^2 (\alpha_0 e^{-i\omega t} a^\dagger_1 + \alpha^*_0 e^{i\omega t} a_1) 
		+ |\alpha_0|^4  |\Psi' \rangle 
		\nonumber \\
		= \dfrac{1}{\mathcal{N}(t)} 
		\bigg[ |\alpha_0|^4 \mathcal{N}(t)  
		+ \sum^{M}_{k=1} |\Theta_k|^2 \big[ k(k-1) + 4|\alpha|^2 k\big] t^4 
		\nonumber \\
		+ \sum^{M}_{k=0}\sum^{M}_{j=n_1}|\Omega_{j,k}|^2 
		\big[ k(k-1) + 4|\alpha_0|^2 k\big] t^6 
		\nonumber \\
		-2|\alpha_0|^2 \big(\Theta_1\alpha^*_{0} + \Theta^*_1\alpha_{0}\big)t^2
		\nonumber 	\\
		+2|\alpha_0|^2 \sum^{M}_{k=1}\sqrt{k+1}
		(\alpha_{0} \Theta^*_{k+1}\Theta_{k} + \alpha^*_{0} \Theta^*_{k}\Theta_{k+1} ) t^4
		\nonumber \\
		+ 2|\alpha_0|^2 \sum_{k=0}^{M} \sum_{j=n_1}^{M}
		\sqrt{k+1}(\alpha_{0} \Omega^*_{j,k+1}\Omega_{j,k} + \alpha^*_{0} \Omega^*_{j,k}\Omega_{j,k+1} ) t^6
		\nonumber \\
		- \sqrt{2} \big(\alpha^{*2}_0 \Theta_2 + \alpha^{2}_0 \Theta^*_2 \big)t^2
		\nonumber \\
		+ \sum^{M}_{k=1}\sqrt{k+1}\sqrt{k+2}
		\big(\alpha^{*2}_{0} \Theta^*_{k}\Theta_{k+2} + \alpha^{2}_{0} \Theta^*_{k+2}\Theta_{k} \big) t^4
		\nonumber\\ 	+
		\sum_{k=0}^{M} \sum_{j=n_1}^{M}
		\sqrt{k+1}\sqrt{k+2} \big(\alpha^{*2}_{0} \Omega^*_{j,k}\Omega_{j,k+2} + \alpha^{2}_{0} \Omega^*_{j,k+2}\Omega_{j,k} \big) t^6
		\nonumber \\
		+2 \sum^{M}_{k=0}k\sqrt{k+1}
		(\alpha_{0} \Theta^*_{k+1}\Theta_{k} + \alpha^*_{0} \Theta^*_{k}\Theta_{k+1} ) t^4 ,
	\end{align}
	and for the $n'$th harmonic mode, $n\neq 1$, it is
	\begin{align}
		G_{nn}(t)= \tfrac{1}{\mathcal{N}(t)} \langle \Psi'| 
		(a^\dagger_n + it\chi_n\alpha^{*n}_{0}e^{in\omega t} )
		(a^\dagger_n + it\chi_n\alpha^{*n}_{0}e^{in\omega t} )
		\nonumber\\
		\times
		(a_n - it\chi_n\alpha^{n}_{0}e^{-in\omega t} )
		(a_n - it\chi_n\alpha^{n}_{0}e^{-in\omega t} )  |\Psi' \rangle
		= \tfrac{1}{\mathcal{N}(t)}
		\nonumber \\
		\times \langle \Psi'| 
		a^{\dagger2}_n a^{2}_n 
		- 2(it\chi_n\alpha^{n}_{0}e^{-in\omega t} a^{\dagger2}_n a_n 
		- it\chi_n\alpha^{*n}_{0}e^{in\omega t} a^{\dagger}_n a^2_n )
		\nonumber\\
		- \chi^2_n t^2 \big(
		\alpha^{2n}_0 e^{-i2n\omega t} a^{\dagger2}_n 
		+ \alpha^{*2n}_0 e^{i2n\omega t} a^{2}_n \big)
		+ 4 \chi^2_n t^2 |\alpha_0|^{2n} a_n^\dagger a_n
		\nonumber\\
		- 2i\chi^3_n |\alpha_0|^{2n} t^3 
		(\alpha^{n}_0 e^{-in\omega t} a^\dagger_n
		-\alpha^{*n}_0 e^{in\omega t} a_n)  
		+ \chi^4_n  t^4 |\alpha_0|^{4n} |\Psi' \rangle 
		\nonumber\\
		= \dfrac{1}{\mathcal{N}(t)} \bigg[
		4\chi^2_n |\alpha_0|^{2n} \sum^{M}_{k=0} |\Omega_{n,k}|^2 t^8
		\nonumber\\	+ 2\chi^3_n |\alpha_0|^{2n} \big(
		\alpha^{*n}\Omega_{n,0} + \alpha^{n}\Omega^*_{n,0} \big) t^6
		+ \chi^4_n  t^4 |\alpha_0|^{4n} \mathcal{N}(t)	
		\nonumber\\ 
		- 2\chi^3_n |\alpha_0|^{2n} \sum^{M}_{k=1}\big(
		\alpha^{*n}\Omega_{n,k}\Theta^*_k + \alpha^{n}\Omega^*_{n,k}\Theta_k
		\big) t^8 \bigg].
	\end{align} 
	Intermodal cross-coherence functions between the excitation mode and the $n'$th harmonic mode is
	\begin{align}
		G_{1n}(t)=
		\dfrac{1}{\mathcal{N}(t)} 
		\langle \Psi' | \big( N_1 + \alpha_{0}e^{-i\omega t}a^\dagger_1 + \alpha^*_{0}e^{i\omega t}a_1 + |\alpha_0|^2 \big)
		\nonumber \\
		\big( N_n - it\chi_n \alpha^n_{0}e^{-in\omega t}a^\dagger_n 
		+ it\chi_n \alpha^{*n}_{0}e^{in\omega t}a_n 
		+ t^2\chi^2_n|\alpha_{0}|^{2n} \big)
		|\Psi' \rangle 
		\nonumber \\ 
		=
		\dfrac{1}{\mathcal{N}(t)} 
		\langle \Psi' | N_1 N_n 
		+ N_n \big( \alpha^{}_{0}e^{-i\omega t} a^\dagger_1 
		+ \alpha^{*}_{0}e^{in\omega t} a_1 + |\alpha_0|^2 \big)
		\nonumber \\
		+ i \chi_n t \big( \alpha^{*n}_{0}e^{in\omega t} N_1 a_n
		-  \alpha^{n}_{0}e^{-in\omega t} N_1 a^{\dagger}_n \big)
		+ \chi^2_n |\alpha_0|^{2n} t^2 N_1
		\nonumber \\
		+ i\chi_n t \big( \alpha^{*(n+1)}_{0}e^{i(n+1)\omega t} a_1 a_n 
		- \alpha^{(n+1)}_{0}e^{-i(n+1)\omega t} a^{\dagger}_1 a^{\dagger}_n 
		\nonumber \\
		+ \alpha^{*n}_{0} \alpha_0 e^{i(n-1)\omega t} a^{\dagger}_1 a_n 
		- \alpha^{n}_{0}\alpha^*_0 e^{-i(n-1)\omega t} a_1 a^{\dagger}_n   \big)
		\nonumber\\
		+ i\chi_n t |\alpha_{0}|^2 \big( \alpha^{*n}_0 e^{in\omega t} a_n
		- \alpha^{n}_0 e^{-in\omega t} a^\dagger_n	\big)
		\nonumber \\ + \chi^2_n |\alpha_{0}|^{2n}  \big(  \alpha_0 e^{-i\omega t} a^\dagger_1 
		+ \alpha^*_0 e^{i\omega t} a_1 
		+ |\alpha_0|^2 \big) t^2
		|\Psi' \rangle
		\nonumber \\
		= \dfrac{1}{\mathcal{N}(t)}\bigg[ 
		\sum_{k=0}^{M} |\Omega_{n,k}|^2 t^6 \big( k
		+ |\alpha_0|^2 \big) 
		\nonumber \\
		+ \sum_{k=0}^{M} \big( \alpha_0 \Omega^*_{n,k+1}\Omega_{n,k} + \alpha^*_0 \Omega^*_{n,k}\Omega_{n,k+1} \big)\sqrt{k+1} t^6
		\nonumber \\
		+ \chi^2_n |\alpha_0|^{2n} t^6
		\bigg( \sum_{k=1}^{M} k|\Theta_{k}|^2 
		+ \sum_{k=0}^{M} \sum_{j=n_1}^{M} k |\Omega_{j,k}|^2 t^2 \bigg)
		\nonumber \\
		-\chi_n t^6 \sum_{k=1}^{M} k 
		\big(\alpha^{*n}_0  \Theta^*_{k}\Omega_{n,k}
		+ \alpha^{n}_0  \Theta_{k}\Omega^*_{n,k} \big)
		\nonumber \\
		+ \chi_n t^4 \bigg(\alpha^{*(n+1)}_0\Omega_{n,1} + \alpha^{(n+1)}_0\Omega^*_{n,1} 
		\nonumber \\		
		- \sum_{k=2}^{M} \sqrt{k} \big( \alpha^{*(n+1)}_0\Theta^*_{k-1} \Omega_{n,k} 
		+ \alpha^{(n+1)}_0\Theta_{k-1} \Omega^*_{n,k} \big) t^2 \bigg)
		\nonumber\\
		- \chi_n |\alpha_0|^2 t^6 
		\nonumber \\
		\times \bigg( 
		\sum_{k=0}^{M} \sqrt{k+1} \big( \alpha^{*(n-1)}_0\Theta^*_{k+1} \Omega_{n,k} 
		+ \alpha^{(n-1)}_0\Theta_{k+1} \Omega^*_{n,k} \big)  \bigg)
		\nonumber
	\end{align}\begin{align}
		+ \chi^2_n |\alpha_{0}|^{2n+2} t^2 \mathcal{N}(t)
		+ \chi_n |\alpha_0|^2 t^4
		\nonumber \\
		\times \bigg(\alpha^{*n}_{0}\Omega_{n,0} + \alpha^{n}_{0}\Omega^*_{n,0} 
		- \sum_{k=1}^{N} \big( \alpha^{*n}_{0} \Theta^*_{k}\Omega_{n,k}
		+ \alpha^{n}_{0} \Theta_{k}\Omega^*_{n,k} \big) t^2 \bigg)
		\nonumber \\
		- \chi^2_n |\alpha_0|^{2n} t^4
		\big(\alpha^{*}_{0}\Theta_{1} + \alpha_{0}\Theta^*_{1} \big) 
		\nonumber\\
		+ \chi^2_n |\alpha_0|^{2n} t^2 \bigg( 
		\sum_{k=1}^{M} \sqrt{k+1} \big( \alpha^{*}_{0}\Theta^*_{k}\Theta_{k+1}
		+ \alpha_{0}\Theta_{k}\Theta^*_{k+1} \big) t^4
		\nonumber\\
		+ \sum_{k=0}^{M}\sum_{j=n_1}^{M} \sqrt{k+1} \big( \alpha^{*}_{0}\Omega^*_{j,k}\Omega_{j,k+1} + \alpha_{0}\Omega_{j,k}\Omega^*_{j,k+1} \big)t^6 
		\bigg) \bigg] ,
	\end{align}
	and between the $n'$th and $m'$th harmonic modes, $n\neq m\neq 1$, it is
	\begin{align}
		G_{nm}(t)= \dfrac{1}{\mathcal{N}(t)}\times
		\nonumber \\
		\langle \Psi' | \big( N_n - it\chi_n (\alpha^n_{0}e^{-in\omega t}a^\dagger_n 
		- \alpha^{*n}_{0}e^{in\omega t}a_n )
		+ t^2\chi^2_n|\alpha_{0}|^{2n} \big) 
		\nonumber \\
		\big( N_m
		- it\chi_m (\alpha^m_{0}e^{-im\omega t}a^\dagger_m 
		- \alpha^{*m}_{0}e^{im\omega t} a_m )
		\nonumber \\
		+ t^2\chi^2_m|\alpha_{0}|^{2m} \big)
		|\Psi' \rangle
		=\dfrac{1}{\mathcal{N}(t)} \langle \Psi|
		N_n N_m 
		\nonumber \\
		-i\chi_n t\big( \alpha^n_0 e^{-in\omega t} a^\dagger_n N_m
		- \alpha^{*n}_0 e^{in\omega t} a_n N_m \big)
		\nonumber \\	
		+ i\chi_m  t \big(\alpha^{*m}_0 e^{im\omega t} a_m N_n 
		- \alpha^m_0 e^{-im\omega t} a^\dagger_m N_n  \big)
		\nonumber 
	\end{align}\pagebreak\begin{align}
		+ N_n \chi^2_m t^2 |\alpha_0|^{2m} + N_m \chi^2_n t^2 |\alpha_0|^{2n}
		\nonumber\\
		- \chi_n \chi_m t^2 \big( \alpha^{n+m}_0 e^{-i(n+m)\omega t} a^\dagger_n a^\dagger_m 
		\nonumber \\
		+ \alpha^{*(n+m)}_0 e^{i(n+m)\omega t} a_n a_m 
		\big)
		\nonumber \\
		+ \chi_n \chi_m t^2 \big( \alpha^{n}_0\alpha^{*m}_0 e^{-i(n-m)\omega t} a^\dagger_n a_m 
		\nonumber \\
		+ \alpha^{*n}_0\alpha^m_0 e^{i(n-m)\omega t} a_n a^\dagger_m 
		\big)
		\nonumber \\
		+ \chi^2_n \chi^2_m t^4 |\alpha_0|^{2n+2m}
		-i\chi_n \chi^2_m |\alpha_0|^{2m} t^3 
		\nonumber \\		\big( e^{-in\omega t}\alpha^{n}_0 a^\dagger_n
		- e^{in\omega t}\alpha^{*n}_0 a_n   \big)
		\nonumber \\
		-i\chi_m \chi^2_n |\alpha_0|^{2n} t^3 
		\big( e^{-im\omega t}\alpha^{m}_0 a^\dagger_m
		- e^{im\omega t}\alpha^{*m}_0 a_m   \big)
		|\Psi \rangle 
		\nonumber \\
		=
		\dfrac{1}{\mathcal{N}(t)} \bigg[ 
		\sum_{k=0}^{M}
		\big[ |\Omega_{n,k}|^2 \chi^2_m |\alpha_0|^{2m} 
		+ |\Omega_{m,k}|^2 \chi^2_n |\alpha_0|^{2n} \big] t^8 
		\nonumber \\ 
		+ \chi_n^2 \chi_m^2 t^4 |\alpha_0|^{2n+2m} \mathcal{N}(t)
		\nonumber \\
		+ \chi_n \chi_m t^8 \sum_{k=0}^{M} 
		\big(\alpha_0^{n}\alpha_0^{*m} \Omega_{m,k}\Omega^*_{n,k}
		+ \alpha_0^{m}\alpha_0^{*n} \Omega_{n,k}\Omega^*_{m,k} \big)
		\nonumber \\+
		\chi_n \chi^2_m t^6 |\alpha_0|^{2m} \big( \alpha_0^{n} \Omega^*_{n,0}
		+ \alpha_0^{*n} \Omega_{n,0} \big)
		\nonumber \\
		+
		\chi^2_n \chi_m t^6 |\alpha_0|^{2n} \big( \alpha_0^{m} \Omega^*_{m,0}
		+ \alpha_0^{*m} \Omega_{m,0} \big)
		\nonumber\\ -
		\chi_n \chi^2_m t^8 |\alpha_0|^{2m} \sum_{k=1}^{M}\big( \alpha_0^{n}\Omega^*_{n,k} \Theta_{k}
		+ \alpha_0^{*n}\Omega_{n,k} \Theta^*_{k} \big)
		\nonumber \\
		-
		\chi_m \chi^2_n t^8 |\alpha_0|^{2n} \sum_{k=1}^{M}\big( \alpha_0^{m}\Omega^*_{m,k} \Theta_{k}
		+ \alpha_0^{*m}\Omega_{m,k} \Theta^*_{k} \big)
		\bigg].
	\end{align}
	\begin{widetext}
		\section{Density matrix of the field}\label{densityM}
		Here, we present the density matrix of the total electromagnetic field $\rho'^{(2)}(t)$, based on the perturbatively calculated quantum state defined by Eq.~(\ref{31egy}), as well as that of various sub-systems that we investigated in our work. Note that here, we let the lowest-order of harmonics be set symbolically to $n_1$ instead of 2. By substituting Eq.~(\ref{31egy}) into $\rho'^{(2)}(t)=|\Psi'(t)\rangle^{(2)} \langle \Psi'(t)|^{(2)}$ and expanding, we get
		\begin{multline}
			\rho'^{(2)}(t) =
			\bigg[ |0 \rangle \!\otimes_n \!
			\vert 0 \rangle_{n}	- \sum_{k=1}^{M} \Theta_k t^2
			| k \rangle  \!\otimes_n\!  \vert 0 \rangle_{n}  e^{-ik\omega t}
			-i	\sum_{k=0}^{M} \sum_{j=n_{1}}^{M}
			\Omega_{j,k} t^3 ~ | k \rangle \vert 1 \rangle_{j}
			\underset{n\neq j}{\otimes} \!\vert 0 \rangle_{n}   
			e^{-i(k+j)\omega t} \bigg]
			\nonumber\\
			\bigg[ \langle0 | \!\otimes_n \!
			\langle 0 \vert_{n}	- \sum_{k=1}^{M} \Theta^*_k t^2
			\langle k |  \!\otimes_n\!  \langle 0 \vert_{n}  e^{ik\omega t}
			+i	\sum_{k=0}^{M} \sum_{j=n_{1}}^{M}
			\Omega^*_{j,k} t^3 ~ \langle k | \langle 1 |_{j}
			\underset{n\neq j}{\otimes} \!\langle 0 |_{n}   
			e^{i(k+j)\omega t} \bigg]
			=
			|0 \rangle\langle 0| \!\otimes_n \! |0 \rangle_{n}\langle 0|_n
			\nonumber \\
			+ 
			\sum_{k,k'=1}^{M} \Theta_k \Theta^*_{k'} t^4 e^{i(k'-k)\omega t}
			| k \rangle\langle k'| \!\otimes_n\! \vert 0 \rangle_{n}\langle 0|_n  
			- \sum_{k=1}^{M} \Theta^*_k t^2 e^{ik\omega t}
			|0 \rangle \langle k | \!\otimes_n \! \vert 0 \rangle_{n} \langle 0 \vert_{n}  
			- \sum_{k=1}^{M} \Theta_k t^2 e^{-ik\omega t} |k \rangle \langle 0| \!\otimes_n \! \vert 0 \rangle_{n} \langle 0 \vert_{n} 
			\nonumber
			\nonumber \\ 
			+ 
			i\sum_{k=0}^{M}\sum_{j=n_1}^{M} \Omega^*_{j,k} t^3 e^{i(k+j)\omega t}
			|0 \rangle \langle k | |0\rangle_j \langle1|_j
			\!\underset{n\neq j}{\otimes}\! \vert 0 \rangle_{n} \langle 0 \vert_{n} 
			- i\sum_{k=0}^{M}\sum_{j=n_1}^{M}\Omega_{j,k} t^3 e^{-i(k+j)\omega t}
			|k \rangle \langle 0 | |1\rangle_j \langle0|_j \!\underset{n\neq j}{\otimes}\! \vert 0 \rangle_{n} \langle 0 \vert_{n}  
			\nonumber \\ +
			\sum_{k,k'=0}^{M} \sum_{j,j'=n_{1}}^{M} \Omega_{j,k}\Omega^*_{j',k'} t^6
			e^{i(k'-k+j'-j)\omega t} 
			~ | k \rangle\langle k'| |1 \rangle_{j}\langle 1|_{j'} \!\!\underset{n\neq j,m\neq j'}{\otimes}\!\! | 0 \rangle_{n}\langle 0|_m 
			\nonumber 
		\end{multline} \begin{multline}
			-i \!\sum_{\substack{k=1\\k'=0}}^{M} \sum_{j=n_{1}}^{M} \Theta_k\Omega^*_{j,k'} t^5
			e^{i(k'-k+j)\omega t} 
			~ | k \rangle\langle k'| |0 \rangle_{j}\langle 1|_{j} \!\underset{n\neq j}{\otimes}\! | 0 \rangle_{n}\langle 0|_n
			+i \!\sum_{\substack{k=1\\k'=0}}^{M} \sum_{j=n_{1}}^{M} \Theta^*_k\Omega_{j,k'} t^5
			e^{-i(k'-k+j)\omega t} 
			~ | k' \rangle\langle k| |1 \rangle_{j}\langle 0|_{j} \!\underset{n\neq j}{\otimes}\! | 0 \rangle_{n}\langle 0|_n ~.
		\end{multline}
		Tracing out the excitation mode leads us to the density matrix associated with the collective harmonic field. Although not written explicitly, the index $n$ runs from $n_1$ to $M$:
		\begin{align}
			\rho'^{(2)}_{\text{harm}}(t) \! \equiv \!
			\sum_i \langle i | \rho'^{(2)}(t) |i\rangle \!=\!
			\nonumber\\
			\bigg[1 \!+\! \sum_{k=1}^{M} |\Theta_k|^2 t^4 \bigg] \!\otimes_n\!  |0 \rangle_{n}\langle 0|_n +
			it^3 \!\! \sum_{j=n_1}^{M} \bigg[ \Omega^*_{j,0} e^{i j\omega t}
			|0\rangle_j \langle1|_j 
			- \Omega_{j,0} e^{-i j\omega t}
			|1\rangle_j \langle0|_j \bigg]
			\!\!\underset{n\neq j}{\otimes} \!\! \vert 0 \rangle_{n} \langle 0 \vert_{n}  
			\nonumber 
		\end{align} \begin{align}
			-i \sum_{k=1}^{M} \sum_{j=n_{1}}^{M} \Theta_k\Omega^*_{j,k} t^5
			e^{ij\omega t} 
			~ |0 \rangle_{j}\langle 1|_{j} \!\!\underset{n\neq j}{\otimes} \!\! | 0 \rangle_{n}\langle 0|_n
			+i \sum_{k=1}^{M} \sum_{j=n_{1}}^{M} \Theta^*_k\Omega_{j,k} t^5
			e^{-ij\omega t} 
			~ |1 \rangle_{j}\langle 0|_{j} \!\!\underset{n\neq j}{\otimes} \!\! | 0 \rangle_{n}\langle 0|_n 
			\nonumber \\ 
			+ \sum_{k=0}^{M} \sum_{j,j'=n_{1}}^{M} \!\! \Omega_{j,k}\Omega^*_{j',k} t^6
			e^{i(j'-j)\omega t} 
			~ |1 \rangle_{j}\langle 1|_{j'} \!\!\underset{n\neq j,m\neq j'}{\otimes} \!\!  | 0 \rangle_{n}\langle 0|_m . 
		\end{align}
		Tracing out all but the $n'$th harmonic mode results in the density matrix associated with the subsystem of a single harmonic mode
		\begin{align}
			\rho'^{(2)}_{\text{n'th}}(t) =
			\sum_{k=0}^{N} |\Omega_{n,k}|^2 t^6 
			~ |1 \rangle_{n}\langle 1|_{n} 
			+ \bigg[1 + \sum_{k=1}^{M} |\Theta_k|^2 t^4 
			+
			\sum_{k=0}^{M} \sum_{\substack{j=n_{1}\\j\neq n}}^{M} |\Omega_{j,k}|^2 t^6 
			\bigg]
			|0 \rangle_{n}\langle 0|_n
			\nonumber\\
			+ 	it^3 \bigg[ \Omega^*_{n,0} e^{in\omega t}
			|0\rangle_n \langle1|_n 
			- \Omega_{n,0} e^{-i n\omega t}
			|1\rangle_n \langle0|_n \bigg]  
			-i t^5 \sum_{k=1}^{M} \bigg[ \Theta_k\Omega^*_{n,k} e^{in\omega t} 
			~ |0 \rangle_{n}\langle 1|_{n} 
			- \Theta^*_k\Omega_{n,k} e^{-in\omega t} 
			~ |1 \rangle_{n}\langle 0|_{n} \bigg]  .
		\end{align}
	\end{widetext}
	To evaluate the entanglement between two harmonic modes, starting from $\rho'^{(2)}_{\text{harm}}(t)$, one needs to trace out all but the n'th and m'th   harmonic modes.
	This leads to
	\begin{align}
		\rho'^{(2)}_{\text{nm}}(t) =
		\bigg[1 + \sum_{k=1}^{M} |\Theta_k|^2 t^4 \bigg] |0 \rangle_{n}\langle 0|_n |0 \rangle_{m}\langle 0|_m 
		\nonumber\\ 
		+ it^3 \bigg[ \Omega^*_{n,0} e^{i n\omega t}
		|0\rangle_n \langle1|_n 
		- \Omega_{n,0} e^{-i n\omega t}
		|1\rangle_n \langle0|_n \bigg]
		\! \vert 0 \rangle_{m} \langle 0 \vert_{m}  
		\nonumber\\
		+ it^3 \bigg[ \Omega^*_{m,0} e^{i m\omega t}
		|0\rangle_m \langle1|_m 
		- \Omega_{m,0} e^{-i m\omega t}
		|1\rangle_m \langle0|_m \bigg]
		\! \vert 0 \rangle_{n} \langle 0 \vert_{n} \nonumber 
		\nonumber \\
		-i \sum_{k=1}^{M} \Theta_k\Omega^*_{n,k} t^5
		e^{in\omega t} 
		~ |0 \rangle_{n}\langle 1|_{n} |0\rangle_{m}\langle 0|_m
		\nonumber\\
		-i \sum_{k=1}^{M} \Theta_k\Omega^*_{m,k} t^5
		e^{im\omega t} 
		~ |0 \rangle_{m}\langle 1|_{m} | 0 \rangle_{n}\langle 0|_n
		\nonumber\\ 
		+i \sum_{k=1}^{M} \Theta^*_k\Omega_{n,k} t^5
		e^{-in\omega t} 
		~ |1 \rangle_{n}\langle 0|_{n} |0 \rangle_{m}\langle 0|_m 
		\nonumber \\ 
		+i \sum_{k=1}^{M} \Theta^*_k\Omega_{m,k} t^5
		e^{-im\omega t} 
		~ |1 \rangle_{m}\langle 0|_{m} |0 \rangle_{n}\langle 0|_n \nonumber
		\nonumber 
	\end{align} \begin{align}
		+ \sum_{k=0}^{M} \Omega_{n,k}\Omega^*_{m,k} t^6
		e^{i(m-n)\omega t} 
		~ |1 \rangle_{n}\langle 1|_{m}  | 0 \rangle_{m}\langle 0|_n 
		\nonumber \\ 
		+ \sum_{k=0}^{M} \Omega_{m,k}\Omega^*_{n,k} t^6
		e^{i(n-m)\omega t} 
		~ |1 \rangle_{m}\langle 1|_{n}  | 0 \rangle_{n}\langle 0|_m 
		\nonumber \\ 
		+ \sum_{k=0}^{M} \Omega_{n,k}\Omega^*_{n,k} t^6 
		~ |1 \rangle_{n}\langle 1|_{n}  | 0 \rangle_{m}\langle 0|_m 
		\nonumber \\ 
		+ \sum_{k=0}^{M} \Omega_{m,k}\Omega^*_{m,k} t^6 
		~ |1 \rangle_{m}\langle 1|_{m}  | 0 \rangle_{n}\langle 0|_n 
		\nonumber \\ 
		+ \sum_{k=0}^{M}\sum_{j=n_{1},j\neq n,m}^{M}\Omega_{j,k}\Omega^*_{j,k} t^6 
		~ |0 \rangle_{n}\langle 0|_{n}  | 0 \rangle_{m}\langle 0|_m ,
	\end{align}
	which, after simplification and by introducing the $|ij\rangle\langle kl|\equiv |i\rangle_n|j\rangle_m\langle k|_n\langle l|_m$ notation, can be rewritten as:
	\begin{align}\label{twomodedensitiy}
		\rho'^{(2)}_{\text{nm}}(t) =
		\bigg[1 + \sum_{k=1}^{M} |\Theta_k|^2 t^4 
		+ \sum_{k=0}^{M}
		\sum_{\substack{j=n_{1}\\j\neq n,m}}^{M} |\Omega_{j,k}|^2 t^6 \bigg] |00 \rangle\langle 00| 
		\nonumber \\ 
		+ it^3 \bigg[ \Omega^*_{n,0} e^{i n\omega t}
		|00 \rangle \langle 10| 
		- \Omega_{n,0} e^{-i n\omega t}
		|10\rangle \langle 00|\bigg]  
		\nonumber
	\end{align} \begin{align}
		+ it^3 \bigg[ \Omega^*_{m,0} e^{i m\omega t}
		|00\rangle \langle 01| 
		- \Omega_{m,0} e^{-i m\omega t}
		|01\rangle \langle 00| \bigg]  
		\nonumber \\
		-i t^5 \sum_{k=1}^{M} \bigg[ \Theta_k\Omega^*_{n,k}
		e^{in\omega t} 
		~ |00 \rangle \langle 1 0|
		+ \Theta_k\Omega^*_{m,k}
		e^{im\omega t} 
		~ |00 \rangle \langle 0 1| \bigg]
		\nonumber \\	
		+i t^5 \sum_{k=1}^{M}\bigg[  \Theta^*_k\Omega_{n,k}
		e^{-in\omega t} 
		~ |10 \rangle \langle 00|
		+  \Theta^*_k\Omega_{m,k}
		e^{-im\omega t} 
		~ |01 \rangle \langle 00|  \bigg]
		\nonumber \\ 
		+ \sum_{k=0}^{M} \Omega_{n,k}\Omega^*_{m,k} t^6
		e^{i(m-n)\omega t} 
		~ |10 \rangle \langle 01| 
		\nonumber \\ 
		+ \sum_{k=0}^{M} \Omega_{m,k}\Omega^*_{n,k} t^6
		e^{i(n-m)\omega t} 
		~ |01 \rangle \langle 10| 
		\nonumber\\ 
		+ \sum_{k=0}^{M} |\Omega_{n,k}|^2 t^6 
		~ |10 \rangle\langle 10| 
		+ \sum_{k=0}^{M} |\Omega_{m,k}|^2 t^6 
		~ |01 \rangle \langle 01|  .
	\end{align}
	
	This, together with its partial transpose with respect to the mode indexed with $n$, can be represented in matrix form as:
	\begin{equation}\label{lognegmatr}
		\rho'^{(2)}_{\text{nm}}(t) =
		\begin{bmatrix}
			a & b & c & 0
			\\
			b^* & d & e & 0
			\\
			c^* & e^* & f & 0
			\\
			0 & 0 & 0 & 0
		\end{bmatrix},
		~~~~~
		\left(\rho'^{(2)}_{\text{nm}}(t)\right)^{\Gamma_n} =
		\begin{bmatrix}
			a & b & c^* & e^*
			\\
			b^* & d & 0 & 0
			\\
			c & 0 & f & 0
			\\
			e & 0 & 0 & 0
		\end{bmatrix},
	\end{equation}
	where 
	\begin{align*}
		a \equiv 1 + \sum_{k=1}^{M} |\Theta_k|^2 t^4 
		+ \sum_{k=0}^{M}
		\sum_{\substack{j=n_{1}\\j\neq n,m}}^{M} |\Omega_{j,k}|^2 t^6 ,
		\\
		b \equiv it^3 \Omega^*_{m,0} e^{im\omega t}
		- i t^5 \sum_{k=1}^{M} \Theta_k \Omega^*_{m,k} e^{im\omega t},
		\\
		c \equiv it^3 \Omega^*_{n,0} e^{in\omega t}
		- i t^5 \sum_{k=1}^{M} \Theta_k \Omega^*_{n,k} e^{in\omega t},
		\\
		d \equiv \sum_{k=0}^{M} |\Omega_{m,k}|^2 t^6,
		~~~
		f\equiv \sum_{k=0}^{M} |\Omega_{n,k}|^2 t^6,
		\\
		e \equiv \sum_{k=0}^{M} \Omega_{m,k}\Omega^*_{n,k} t^6 e^{i(n-m)\omega t}.
	\end{align*}
	
	For the sake of completeness, let us remind the reader that when determining the Wigner-function from a density-operator, the decomposition $\rho=\sum_{m,n}\rho_{m,n}|m\rangle\langle n|$ can be used, with which the Wigner-function can be written as 
	\begin{equation}
		W(q,p)=\sum_{m,n}\rho_{m,n} W_{mn}(q,p) .
	\end{equation} 
	Here
	\begin{equation}
		W_{mn}(q,p) = \dfrac{(-1)^m}{\pi} \langle n| \mathcal{D}(2\alpha) |m\rangle,
	\end{equation}
	and 
	\begin{align}
		\langle n| \mathcal{D}(2\alpha) |m\rangle
		= \nonumber\\
		\begin{Bmatrix}
			\sqrt{\frac{n!}{m!}} e^{-2|\alpha|^2}(-\alpha^*)^{m-n}L^{m-n}_n(4|\alpha|^2) & ~~m\geq n
			\\
			\sqrt{\frac{m!}{n!}}
			e^{-2|\alpha|^2}(\alpha)^{n-m}L^{n-m}_m(4|\alpha|^2) & ~~m< n
		\end{Bmatrix}.
	\end{align}

	\section{Validity of the perturbative expansion}\label{addendumvalidity}
	
		Choosing the magnitude of the largest $\chi_n t$ product as the perturbative parameter, we estimate both the error introduced by truncating the perturbative expansion and the condition required for its convergence. Rearranging the quantum state into terms $|\mathcal{O}(\chi^j t^j)\rangle$, defined to be proportional to $\chi^j t^j$, we calculate the leading-order neglected term, $|\mathcal{O}(\chi^4 t^4)\rangle$.
		To identify the relevant contributions, recall that the interaction Hamiltonian $W'(t)$ in Eq.~(\ref{whamiltonian}) decomposes as $W'(t) = H'_{DH}(t) + H'_{D}(t)$, with
		\begin{align*}
			H'_{DH}(t) =
			\hbar \sum^M_{n=2}  \chi_n 
			\bigg[ a_n^\dagger 
			\sum_{k=1}^{n} \binom{n}{k} A^{k}  \alpha^{n-k}_0 e^{-i(n-k)\omega t} 
			\nonumber\\ + a_n 
			\sum_{k=1}^{n} \binom{n}{k} A^{\dagger k}  \alpha^{*(n-k)}_0 e^{i(n-k)\omega t}
			\bigg],
			\\ 
			H'_{D}(t) =
			\hbar \sum^M_{n=2} \chi^2_n t
			\bigg[ i \alpha^{*n}_0  
			\sum_{k=0}^{n} \binom{n}{k} A^{k}  \alpha^{n-k}_0 e^{i k\omega t}
			\nonumber\\ - i \alpha^n_0 
			\sum_{k=0}^{n} \binom{n}{k} A^{\dagger k}  \alpha^{*(n-k)}_0 e^{-i k\omega t}
			\bigg]. \label{eq:HD}
		\end{align*}
		$H'_{DH}(t)$ and $H'_{D}(t)$ are proportional to $\chi_n$ and $\chi^2_n$, respectively. Consequently, upon integration of Eq.~(\ref{dynequ}), their action on $|\mathcal{O}(\chi^j t^j)\rangle$ raises the order of $\chi_n t$ by one and by two, respectively. Therefore, identifying the contributions to $|\mathcal{O}(\chi^4 t^4) \rangle$ is straightforward.

	The first contribution consists of the terms neglected in Eq.~(\ref{dynequ2order}), arising from $H'_D(t)$ acting on the second-order term,
		\begin{align}
			H'_D(t) \left[ 
			-\sum_{k'=1}^{M} \Theta_{k'} t^2 |k',0\dots0\rangle
			\right]
			\nonumber \\ = 
			- i\sum_{k'=1}^{M} \sum^M_{n=2}
			\hbar \Theta_{k'} \chi^2_n t^3
			\bigg[  \alpha^{*n}_0  
			\sum_{k=0}^{n} \binom{n}{k} A^{k}  \alpha^{n-k}_0 e^{i k\omega t}
			\nonumber \\
			- \alpha^n_0 
			\sum_{k=0}^{n} \binom{n}{k} A^{\dagger k}  \alpha^{*(n-k)}_0 e^{-i k\omega t}
			\bigg] |k',0\dots0\rangle,
		\end{align}
		which, using the operator $A^{(\dagger)}$ and after integration, yields
		\pagebreak
		\begin{align}
			|\mathcal{O}(\chi^4 t^4)\rangle_{D}= - \sum^M_{n=2} \sum_{k'=1}^M \sum_{k=0}^{\min[n,k']}
			\chi^2_n  \dfrac{t^4}{4}
			\alpha^{*n}_0   \binom{n}{k} 
			\nonumber \\
			\times\sqrt{\dfrac{k'!}{(k'-k)!}}  \alpha^{n-k}_0 
			\Theta_{k'}  |k'-k,0\dots 0\rangle
			\nonumber \\
			+\sum^M_{n=2} \sum_{k'=1}^M \sum_{k=0}^{n}
			\chi^2_n  \dfrac{t^4}{4}
			\alpha^n_0 \binom{n}{k} 
			\nonumber \\
			\times\sqrt{\dfrac{(k'+k)!}{k'!}} 
			\alpha^{*(n-k)}_0 
			\Theta_{k'}  |k'+k,0\dots 0\rangle. \label{eq:term-theta}
		\end{align}

		The second, and only remaining, contribution proportional to $\chi_n^4 t^4$ arises from $H'_{DH}(t)$ acting on the third-order term,
		\begin{align}
			H'_{DH}(t) \left[ 
			-i \sum_{k'=0}^{M} \sum_{j=2}^{M} \Omega_{j,k'} t^3 |k',0\dots 1_j \dots0\rangle
			\right]
			\nonumber\\ = 
			-i \hbar 
			\sum_{k'=0}^{M} \sum_{j=2}^{M}  \sum^M_{n=2} 
			\Omega_{j,k'} t^3  \chi_n 
			\bigg[ a_n^\dagger 
			\sum_{k=1}^{n} \binom{n}{k} A^{k}  \alpha^{n-k}_0 e^{-i(n-k)\omega t} 
			\nonumber\\
			+ a_n 
			\sum_{k=1}^{n} \binom{n}{k} A^{\dagger k}  \alpha^{*(n-k)}_0 e^{i(n-k)\omega t}
			\bigg] |k',0\dots 1_j \dots0\rangle,
		\end{align}
		which, by an analogous calculation, yields
		\begin{align}
			|\mathcal{O}(\chi^4 t^4)\rangle_{DH} =
			- \sum^M_{n=2} \sum_{k'=1}^M  \sum_{k=1}^{\min[n,k']}\sum_{j=2,j\neq n}^M 
			\chi_n \binom{n}{k} \alpha^{n-k}_0  
			\nonumber\\
			\times\Omega_{j,k'} \dfrac{t^4}{4} 
			\sqrt{\dfrac{k'!}{(k'-k)!}} |k'-k,0\dots 1_j\dots 1_n\dots 0\rangle 
			\nonumber\\
			- \sum^M_{n=2} \sum_{k'=1}^M \sum_{k=1}^{\min[n,k']}  
			\chi_n \binom{n}{k}  \alpha^{n-k}_0  
			\Omega_{n,k'} \dfrac{t^4}{4} 
			\nonumber\\
			\times \sqrt{2\dfrac{k'!}{(k'-k)!}} |k'-k,0\dots 2_n\dots 0\rangle 
			\nonumber\\
			-  \sum^M_{n=2} 
			\sum_{k'=0}^M \sum_{k=1}^{n} 
			\chi_n  \binom{n}{k} \sqrt{\frac{(k'+k)!}{k'!}}
			\alpha^{*(n-k)}_0 
			\nonumber\\
			\times\Omega_{n,k'} \dfrac{t^4}{4} |k'+k,0\dots 0\rangle. \label{eq:term-omega}
		\end{align} 
		Since Eqs.~(\ref{eq:term-theta}) and (\ref{eq:term-omega}) exhaust every contribution proportional to $\chi_n^4t^4$, the leading-order neglected term is simply their sum,
		\begin{align}
			|\mathcal{O}(\chi^4 t^4)\rangle = |\mathcal{O}(\chi^4 t^4)\rangle_{D} + |\mathcal{O}(\chi^4 t^4)\rangle_{DH}. \label{eq:chi4t4}
	\end{align} 
	
		Our aim here is to give only a qualitative estimate of the magnitude of this term. We use the norm
		\begin{equation}\label{norma}
			\big|\big| |\mathcal{O}(\chi^j t^j)\rangle \big|\big|= \sqrt{\langle \mathcal{O}(\chi^j t^j) | \mathcal{O}(\chi^j t^j)\rangle}
		\end{equation} 
		to evaluate the (non-normalized) terms of the expansion. To proceed, we replace all $\chi_n$ by the largest value $\chi$, retain only the highest power of $\alpha_0$ appearing in each sum, extend all sums to run over all $M$ modes, and neglect the combinatorial coefficients. With this rule of thumb, we obtain $|\Theta|\sim M \chi^2 |\alpha_0|^{2M}$ and $|\Omega|\sim M^2 \chi^3 |\alpha_0|^{3M}$.

		With these considerations, and referring to Eq.~(\ref{31egy}), the magnitudes of $|\mathcal{O}(\chi^0 t^0)\rangle$, $|\mathcal{O}(\chi^1 t^1)\rangle$, $|\mathcal{O}(\chi^2 t^2)\rangle$, and $|\mathcal{O}(\chi^3 t^3)\rangle$ follow directly. The only zeroth-order term has unit coefficient, and no first-order term exists. There are $M$ orthogonal second-order terms, each of magnitude estimated as $|\Theta|^2t^4$, resulting in $\big|\big| |\mathcal{O}(\chi^2 t^2)\rangle \big|\big|\sim \sqrt{M}|\Theta| t^2$ overall. There are approximately $M^2$ orthogonal third-order terms, and an analogous argument yields $\big|\big| |\mathcal{O}(\chi^3 t^3)\rangle \big|\big|\sim M|\Omega|t^3$.
		Evaluating the fourth-order term requires more care. Applying the same approximations results in
		\begin{align}
			|\mathcal{O}(\chi^4 t^4)\rangle \sim
			- M \sum_{k'=1}^M \sum_{k=0}^{k'}
			\chi^2 t^4
			|\alpha_0|^{2M} 
			\Theta  |k'-k,0\dots 0\rangle
			\nonumber \\
			+M \sum_{k'=1}^M \sum_{k=0}^{M}
			\left(\chi^2 |\alpha_0|^{2M} \Theta 
			-  \chi \alpha^{*M}_0  \Omega \right)t^4
			|k'+k,0\dots 0\rangle   
			\nonumber\\
			- \sum_{\substack{n,j=2\\j\neq n}}^M \sum_{k'=1}^M  \sum_{k=1}^{k'} 
			\chi t^4 \alpha^{M}_0  
			\Omega |k'-k,0\dots 1_j\dots 1_n\dots 0\rangle 
			\nonumber\\
			- \sum_{n=2}^M \sum_{k'=1}^M \sum_{k=1}^{k'}  
			\chi t^4 \alpha^{M}_0  
			\Omega  |k'-k,0\dots 2_n\dots 0\rangle .
		\end{align}
		Rewriting the summations equivalently and separating orthogonal terms yields
		\begin{align}\label{estimatesep}
			|\mathcal{O}(\chi^4 t^4)\rangle \sim
			-M^2 \chi^2 t^4 |\alpha_0|^{2M} \Theta 
			|0,0\dots 0 \rangle
			\nonumber\\
			+ M t^4 \sum_{q=1}^M \left[ (2q-M-1)\chi^2  |\alpha_0|^{2M}  \Theta  
			- q\chi \alpha_0^{*M} \Omega \right]
			|q,0\dots 0\rangle
			\nonumber \\
			+M t^4 \!\!\! \sum_{q=M+1}^{2M}
			(2M-q+1)\left( \chi^2 |\alpha_0|^{2M} \Theta 
			-  \chi \alpha^{*M}_0  \Omega \right)
			|q,0\dots 0\rangle   
			\nonumber\\
			- \chi t^4 \alpha^{M}_0 \Omega 
			\sum_{\substack{n,j=2\\j\neq n}}^M 
			\sum_{q=0}^{M-1} (M-q)
			|q,0\dots 1_j\dots 1_n\dots 0\rangle 
			\nonumber\\
			- \chi t^4 \alpha^{M}_0 \Omega 
			\sum_{n=2}^M \sum_{q=0}^{M-1} (M-q)  
			|q,0\dots 2_n\dots 0\rangle .
		\end{align}
		To proceed, we make use of an upper bound for the resulting coefficients in Eq.~(\ref{estimatesep}), resulting in
		\begin{align}
			|C_{0,\dots0}|^2 \sim
			M^6 \chi^8 t^8 |\alpha_0|^{8M}  
			\nonumber\\
			|C_{q,\dots0}|^2 \sim M^8 \chi^8 t^8 |\alpha_0|^{8M} 
			\nonumber\\
			|C_{q,0\dots1_j\dots1_n\dots0}|^2 \sim M^6 \chi^8 t^8 |\alpha_0|^{8M} 
			\nonumber\\
			|C_{q,0\dots2_n\dots0}|^2 \sim M^6 \chi^8 t^8 |\alpha_0|^{8M} .
		\end{align}
	
		\begin{widetext}
			The corresponding magnitude estimate then follows directly from Eq.~(\ref{norma}) as
			\begin{align}
				\big|\big| |\mathcal{O}(\chi^4 t^4) \rangle \big|\big|
				\sim 
				\bigg( |C_{0,\dots0}|^2 + 2M|C_{q,0\dots0}|^2 
				+ \tfrac{M(M-1)(M-2)}{2}|C_{q,0\dots1_j\dots1_n\dots0}|^2 
				+ M(M-1) |C_{q,0\dots2_n\dots0}|^2
				\bigg)^{1/2} ,
			\end{align}
		\end{widetext}
		where the dominant contribution scales as $M^{9/2}\chi^4 t^4 |\alpha_0|^{4M}$. Collecting the magnitudes at successive orders, we obtain
		\begin{align}
			\big| |\mathcal{O}(\chi^0 t^0) \rangle \big| &\sim 1, 
			\nonumber \\
			\big| |\mathcal{O}(\chi^1 t^1) \rangle \big| &\sim 0,
			\nonumber \\
			\big| |\mathcal{O}(\chi^2 t^2) \rangle \big|
			&\sim M^{3/2} \chi^2 t^2 |\alpha_0|^{2M} , 
			\nonumber \\
			\big| |\mathcal{O}(\chi^3 t^3) \rangle \big|
			&\sim M^3 \chi^3  t^3 |\alpha_0|^{3M} ,
			\nonumber \\
			\big| |\mathcal{O}(\chi^4 t^4) \rangle \big|
			&\sim 
			M^{9/2} \chi^4 t^4 |\alpha_0|^{4M} .
		\end{align}

		Accordingly, the magnitude of the leading-order neglected term can be estimated as $M^{9/2}|\alpha_0|^{4M} \chi^4 t^4$, where $M$ denotes the number of significantly populated harmonic modes. Recalling that $\chi^2 |\alpha_0|^{2M} t^2$ provides an estimate of the mean photon number $\langle N_h\rangle$ of a populated harmonic mode within the zeroth-order approximation of $\langle N_h\rangle$ presented in Eq.~(\ref{meanphotonnharm}), the leading-order correction can be written compactly as $M^{9/2} \langle N_h\rangle^2$. Likewise, the ratio of successive higher-order corrections is $M^{3/2} \chi t |\alpha_0|^{M}$, which can be interpreted physically as $M^{3/2}\langle N_h \rangle^{1/2}$.

	\bibliography{allthebibGA}
	
\end{document}